\documentclass{sig-alternate}

 \pdfoutput=1 

\usepackage{graphicx}
\usepackage{color}
\usepackage{url}
\usepackage{subcaption}
\usepackage{bbm}
\usepackage{epsfig}
\usepackage{epstopdf}
\usepackage{balance} 

\allowdisplaybreaks

\newtheorem{theorem}{Theorem}
\newtheorem{lemma}{Lemma}

\newtheorem{observation}{Observation}

\newtheorem{fact}{Fact}

\newcommand{\eps}{\epsilon}
\newcommand{\abs}[1]{\left| #1 \right|}

\newcommand{\DISJ}{{$s$-DISJ}}
\newcommand{\TDISJ}{{$2$-DISJ}}

\newcommand{\G}{\mathcal{G}}

\newcommand{\ID}{\mathsf{ID}}

\newcommand{\air}{{\tt Airline}}
\newcommand{\house}{{\tt Household}}
\newcommand{\cover}{{\tt Covertype}}
\newcommand{\ind}{{\tt INDI}}
\newcommand{\cor}{{\tt CORR}}
\newcommand{\anti}{{\tt ANTI}}

\newcommand{\naive}{{\tt Naive}}
\newcommand{\opt}{{\tt Optimal}}
\newcommand{\tradeoff}{{\tt Tradeoff}}
\newcommand{\agids}{{\tt AGiDS}}
\newcommand{\fds}{{\tt FDS}}

\newcommand{\prune}{{\tt Prune}}
\newcommand{\bds}{{\tt BDS}}
\newcommand{\ids}{{\tt IDS}}
\newcommand{\pds}{{\tt PDS}}

\renewcommand{\paragraph}[1]{\medskip \noindent {\bf #1.}}

{\unskip\nobreak\hskip 1em plus 1fil\nobreak$\Box$
\parfillskip=0pt%
\endtrivlist}

\usepackage{algpseudocode}
\usepackage{algorithmicx}
\usepackage{algorithm}

\begin{document}
\title{Computing Skylines on Distributed Data}

\numberofauthors{2} 

\author{
\alignauthor
Haoyu Zhang\\
       \affaddr{Indiana University Bloomington}\\
       \affaddr{Bloomington,IN,USA}\\
       \email{hz30@umail.iu.edu}
\alignauthor
Qin Zhang\\
       \affaddr{Indiana University Bloomington}\\
       \affaddr{Bloomington,IN,USA}\\
       \email{qzhangcs@indiana.edu}
}

\date{}

\maketitle

\begin{abstract}
In this paper we study skyline queries in the distributed computational model, where we have $s$ remote sites and a central coordinator (the query node); each site holds a piece of data, and the coordinator wants to compute the skyline of the union of the $s$ datasets. The computation is in terms of rounds, and the goal is to minimize both the total communication cost and the round cost.

Viewing data objects as points in the Euclidean space, we consider both the horizontal data partition case where each site holds a subset of points, and the vertical data partition case where each site holds one coordinate of all the points.  We give a set of algorithms that have provable theoretical guarantees, and complement them with information theoretical lower bounds.  We also demonstrate the superiority of our algorithms over existing heuristics by an extensive set of experiments on both synthetic and real world datasets.
\end{abstract}

\section{Introduction}
\label{sec:intro}

Skyline computation, also known as the {\em maximal vector problem}, is a useful database query for multi-criteria decision making.  If we view data objects as points in the Euclidean space, then the skyline is defined to be the subset of points that cannot be dominated by others, where we say a point $y$ dominates a point $x$ if $y$ dominates $x$ in all dimensions.  This problem was first studied in computational geometry in the mid-1970's~\cite{KLP75}, and was later introduced into databases as a query operator~\cite{BKS01}.  There is a vast literature on skyline computation and its variants, and we refer readers to \cite{CCM13} and \cite{HV12} for excellent surveys.

Most work on skyline computation in the literature has been conducted in the RAM (signal machine) model.  In recently years, due to the large size of the datasets and the popularity of {\em map-reduce} type of computation, a number of parallel skyline algorithms have been proposed~\cite{WZFZAA06,RVDN09,WOTX07,VDK08,KYZ11,AKSU15,MPLZ14,PD15}.  A common feature of these parallel algorithms is that they use the {\em divide-and-conquer} approach, that is, they use central mechanisms to partition the whole point set into a number of subsets, and then assign each subset to a machine for local processing; finally the local results are merged to form the global skyline.  The art of the algorithm design in this line of work lies in how to choose the partition mechanism.  

In this paper we study the skyline computation on distributed data, which is different from parallel computation in that the data is inherently distributed in different locations, and we {\em cannot} afford to repartition the whole dataset since data repartition is  communication prohibited over networks, and may also cause local storage and data privacy issues which the query node cannot control.

Consider a global hotel search engine, where each hotel is represented as a point in the $2$-dimensional Euclidean space with the $x$-coordinate standing for the price and the $y$-coordinate standing for the rate of the location.  A user naturally wants to find a hotel with the best location {\em and} the best price, although in reality hotels in good locations usually have higher prices.  Thus a good search engine should recommend the user with a list of candidates such that no other hotel has both cheaper price  and better location.  This list is exactly the skyline of the point set.   Given a query, the search engine (e.g., {\em kayak.com},  {\em hipmunk.com}) needs to contact servers/providers in different locations worldwide. The total bits of communication between the query node and servers and the communication rounds typically dominate the engine's response time, since sending messages through network is much slower than local computation, and the initialization of a new communication round takes quite some system overhead.

\begin{figure}[t]
\centering
\includegraphics[height = 1.1in]{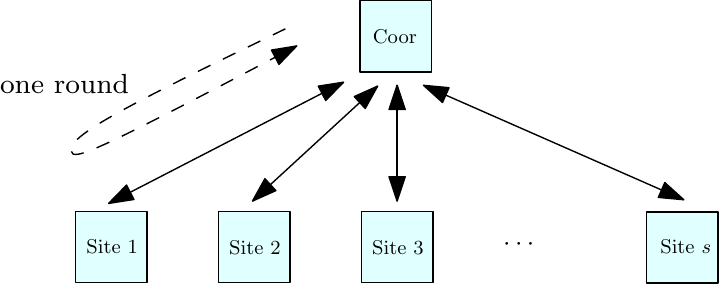}
\caption{The Coordinator Model}
\label{fig:model}
\end{figure}

\paragraph{The Coordinator Model and Previous Work}  We study the skyline problem in the {\em coordinator model} which captures the type of distributed computation mentioned above.   In this model we have $s$ remote sites each holding a piece of data, and a central coordinator which acts as the query node. We assume there is a two-way communication channel between each site and the coordinator.  The computation is in terms of rounds: at the beginning of each round the coordinator sends a message to some of the sites, and then each of the contacted site sends a response back to the coordinator.  The goal is to minimize the total bits of communication and the number of the rounds of the computation.  See Figure~\ref{fig:model} for a visualization of the model. 

We differentiate two scenarios of data storage at sites:  {\em horizontal partition} and {\em vertical partition}.  In the former each site contains an (arbitrary) subset of points. And in the later each site contains {\em one} attribute of {\em all} points together with their IDs which are used to restore the whole point vectors.  Thus if points have $d$ dimensions, we have $d$ sites.  The vertical partition corresponds to the settings where the information of each attribute has to be retrieved from a different server/provider. For example, we may want to find the lowest prices of hotels from {\em kayak.com}, while the customers' best ratings from {\em TripAdvisor}. 

The coordinator model is equivalent to the models used in several previous works for distributed skyline computation \cite{RVDN09,ZTZ09} (horizontal partition) and \cite{BGZ04,LYLC06} (vertical partition).  However, all the existing skyline algorithms in this model are heuristic in nature. We will briefly describe these algorithms in Section~\ref{sec:exp-hor} (horizontal partition) and Section~\ref{sec:heuristic} (vertical partition) respectively. In \cite{TBPY13} the authors studied a general version of vertical partition where multiple columns can be stored in one site. In this paper we will give algorithms with provable theoretical guarantees (for vertical partition we give a better heuristic, given a strong impossibility result that we will show).  We will also compare our algorithms with previous ones experimentally in Section~\ref{sec:exp}.

\paragraph{The Skyline Problem}  
We now give the formal definition of the skyline problem. In this paper we consider the problem in the $2$-dimensional Euclidean space. Given two distinct points $u = (x_u, y_u)$ and $v = (x_v, y_v)$, we say $u$ {\em dominates} $v$, denoted by $u \succeq v$, if $x_u \ge x_v$ {\em and} $y_u \ge y_v$.  For a set of distinct points $S$, the skyline of $S$ is defined to be
$$ sk(S) = \{u \in S\ |\ v \in S, \text{ if }  v \succeq u \text{ then } v = u\}.
$$

We next define the data partition of the skyline problem in the coordinator model. In the horizontal partition case, each site $i$ holds a set $S_i \subseteq {S}$.  In the vertical partition case, we have two sites; the first site holds $\{(x_u, \ID_u)\ |\ u \in S\}$ where $\ID_u$ denotes the ID (key) of point $u$, and the second site holds $\{(\ID_u, y_u)\ |\ u \in S\}$.  In both cases, the coordinator needs to output $sk(S)$ at the end of the computation.

\paragraph{Our Contribution}
We have made the following contributions
in this paper.  Let $n$ be the total number of points, $k$ be the number of skyline points (i.e., the output size), and $s$ be the number of sites. 
\begin{enumerate}
\item  For the horizontal partition, we propose two algorithms. The first one achieves the theoretically {\em optimal} communication cost $\Theta(ks)$, but needs $\lceil k/2 \rceil$ communication rounds.  The second one gives a tradeoff between the total communication cost and rounds: given a round budget $r$ the algorithm uses roughly $O(skn^{2/r})$
communication (see Theorem~\ref{thm:tradeoff} for the precise bound).  
We also prove that any one-round algorithm needs linear (i.e., $\Omega(n)$) communication in the worst case.  These are presented in Section~\ref{sec:horizon}.

\item For the vertical partition, we first prove that in the worst case, any algorithm needs linear communication, regardless of the communication rounds.   We then propose a heuristic that works well in practical settings.  These are presented in Section~\ref{sec:vertical}.

\item We have implemented our algorithms and relevant heuristic  algorithms in the literature, and run them on both synthetic and real-world datasets.  Our experiments have demonstrated the superiority of our algorithms over the existing ones in various aspects.  We also noticed that  for the horizontal partition and the datasets we have tested, with only three communication rounds the tradeoff algorithm can achieve similar communication cost compared with the theoretically optimal one.   The experimental results are presented in Section~\ref{sec:exp}.
\end{enumerate}

\paragraph{Preliminaries}
Let $[n]$ denote $\{1, 2, \ldots, n\}$.  

The $\phi$-quantile of a set $S$ is an element $x$ such that at most $\phi \abs{S}$ elements of $S$ are smaller than $x$ and at most $(1 - \phi)\abs{S}$ elements of $S$ are greater than $x$.  If an $\eps$-approximation is allowed (denoted by $(\phi, \eps)$-quantile), then we can return any $\phi'$-quantile of $S$ such that $\phi - \eps \le \phi' \le \phi$.

%


\section{Horizontal Partition}
\label{sec:horizon}

In this section we give two algorithms for skyline computation in the horizontal partition case. 
We then complement them by proving that the first algorithm is {\em optimal} in terms of the total communication cost (though it may need a large number of rounds). On the other extreme, if we want the computation to be done in one round then in the worst case the sites need to send almost everything to the coordinator.  Our second algorithm gives a tradeoff in between.  

We also show that if data is sorted among the sites according to one of the coordinates, then there is a simple algorithm that has much smaller communication cost and uses only two rounds.

\subsection{Algorithms}
There is a simple algorithm for computing the skyline points in the coordinator model in {\em one} round:  Each site computes the skyline of its local data points and sends it to the coordinator, and then the coordinator computes the global skyline on top of the $s$ local skylines. Unfortunately this algorithm has communication cost $\Omega(n)$; in other words, in the worse case almost all points in all sites need to be sent to the coordinator.  We enclose a proof for this statement in Section~\ref{sec:lb-one}.  We thus try to explore if more rounds can help to reduce the communication cost.

\subsubsection{Algorithm with Optimal Communication Cost}
\label{sec:optimal}

Our algorithm with communication cost $O(ks)$ is described in Algorithm~\ref{alg:optimal}.  We will show  in Section~\ref{sec:lb-infinite} that this communication cost is in fact optimal\footnote{Up to a logarithmic factor which counts the number of bits used to represent a point.}  even if we allow an {\em infinite} number of communication rounds.

\begin{algorithm}[t]
\caption{Optimal Communication under Horizontal Partition}
\label{alg:optimal}
\begin{algorithmic}[1]
\Require $S_i$ is initialized as the point set held by Site $i$
\Ensure the global skyline

\State Site $i$ computes its local skyline points and discards the other points  in $S_i$

\While{$\exists i \in [s]\ s.t.\ S_i \neq \emptyset$}

\For{each $i\ s.t. \ S_{i} \neq \emptyset$}  \label{ln:optimal-1}

\State Site $i$ sends the point with the largest $x$-coordinate and the point with the largest $y$-coordinate to the coordinator   \Comment the two points can be the same point

\EndFor \label{ln:optimal-2}

\State The coordinator computes new global skyline points from the points received from all sites, and sends the new global skyline points to each site \label{ln:optimal-3}

\For{each $i\ s.t. \ S_{i} \neq \emptyset$} 

\State Site $i$ prunes $S_i$ by new global skyline points received from the coordinator  \label{ln:optimal-5}

\EndFor \label{ln:optimal-4}

\EndWhile 
\end{algorithmic}
\end{algorithm}

Let us explain Algorithm~\ref{alg:optimal} in words.  At the beginning, each site computes its local skyline points since only these points can possibly be the global skyline points.  The rest of the algorithm works as follows. At each round, the coordinator tries to find the (at most two) points with the largest $x$-coordinate or $y$-coordinate in the remaining points held by all sites.  This is done by asking all sites to report the local maximums of their remaining points (Line~\ref{ln:optimal-1}-\ref{ln:optimal-2}).  Next, the coordinator computes new global skyline points from the received local maximums, 
and then sends the new global skyline points to all sites for another local pruning step (Line~\ref{ln:optimal-3}-\ref{ln:optimal-4}).

We now show the correctness of Algorithm~\ref{alg:optimal} and analyze its costs.  First, it is clear that the points with the largest $x$-coordinate or $y$-coordinate must be on the skyline.  After the pruning, the points with the largest $x$-coordinate or $y$-coordinate in the remaining points also must be on the skyline since they cannot be dominated by the other remaining points as well as the previous skyline points.  We thus can find one or two skyline points in each round (finding one skyline point may only happen in the last round).  Since there are at most $k$ skyline points, the algorithm will terminate after at most $\lceil k/2 \rceil$ rounds.   

The running time at each site consists of two parts: the computation of the local skylines and the point prunings.  The computation of local skyline at the $i$-th site needs $O(n_i \log n_i)$ time~\cite{KLP75}, where $n_i = \abs{S_i}$ is the number of points at the $i$-th site. The time used for pruning is linear in $n_i$ at the $i$-th site: when pruning points using the new skyline point with the global maximum $y$-coordinate (Line~\ref{ln:optimal-5}), we scan the points which are sorted increasingly according to their $x$-coordinates after the local skyline computation. If a point cannot be pruned in a particular round, then all the points after it with larger $x$-coordinates cannot be pruned in this round. Therefore every point will only be pruned once in the whole computation.  Same arguments hold for the prunings using the new skyline point with the global maximum $x$-coordinate. At the coordinator, for each round we only need to compute the maximums over at most $2s$ points. Thus the total running time is bounded by $O(ks)$.


\begin{theorem}
\label{thm:optimal}

There exists an algorithm for computing the skyline on $n$ points in the $2$-dimensional Euclidean space in the coordinator model with $s$ sites and horizontal data partition that uses $O(ks)$ communication and $\lceil k/2 \rceil$ rounds, where $k$ is the output size, that is, the number of points in the skyline.  The total running time at the $i$-th site is $O(n_i \log n_i)$ where $n_i$ is the number of points at the $i$-th site, and that at the coordinator is $O(ks)$.
\end{theorem}

\subsubsection{A Communication-Round Tradeoff} 
\label{sec:tradeoff}

In the previous section we have shown an algorithm with the optimal communication cost but needs $\lceil k/2 \rceil$ rounds; on the other hand, there is a naive one-round algorithm but in the worst case it needs $\Omega(n)$ bits of communication.  The natural questions is: 
\begin{quote}
{\em Can we obtain a communication-round tradeoff to bridge the two extremes?}  
\end{quote}
We try to address this question by proposing an algorithm that allows the users to choose the number of the rounds of the communication in the computation. In this section we show the following result.   

\begin{theorem}
\label{thm:tradeoff}

There exists an algorithm for computing the skyline on $n$ points in the $2$-dimensional Euclidean space in the coordinator model with $s$ sites and horizontal data partition that uses $r\ (\ge 3)$ rounds and $C = O(sk (n/s)^{1/\lceil r/2 \rceil})$ communication, where $k$ is the output size, that is, the number of points in the skyline.  The total running time at the $i$-th site is $O(C/s + n_i \log n_i)$ where $n_i$ is the number of points at the $i$-th site, and the total running time at the coordinator is $O(C)$.
\end{theorem}

\begin{figure}[t]
\centering
\includegraphics[height = 2in]{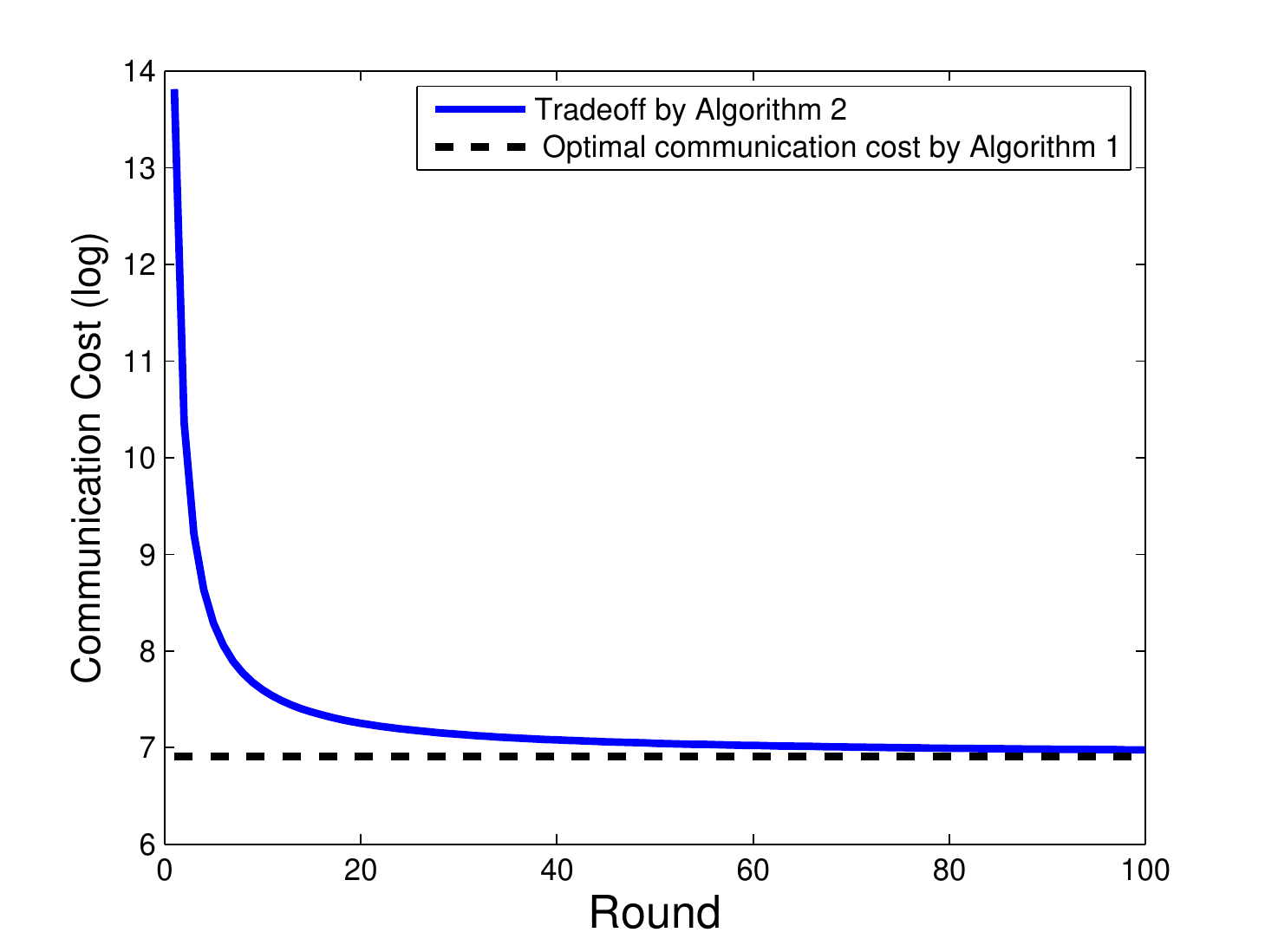}
\caption{Theoretical Communication-Round Tradeoff}
\label{fig:tradeoff}
\end{figure}

Figure~\ref{fig:tradeoff} visualizes the communication-round tradeoff.
\smallskip

Let $t = \lceil r/2 \rceil$.  We describe our tradeoff algorithm in Algorithm~\ref{alg:tradeoff}.  The algorithm again starts with a local skyline computation at each site.  Similar to Algorithm~\ref{alg:optimal}, the rest of the tradeoff algorithm still proceeds in rounds. The main difference is that at each round, the parties (sites and the coordinator) first jointly compute $(1/d, 1/(2d))$-quantiles to partition the Euclidean plane to a set of at most $d$ vertical strips, and then instead of computing the (at most $2$) points with the global maximum $x/y$ coordinates, the coordinator computes for each non-empty strip the point with the maximum $y$-coordinate by collecting information from the sites (Line~\ref{ln:tradeoff-1}-\ref{ln:tradeoff-2}); after that the parties use these points to compute new skyline points and prune each strip.   We call the combination of computing the quantiles and maximum $y$-coordinates, and finding new skyline points and performing local pruning, one {\em step} of the computation.  
The algorithm runs for $(t-1)$ steps, and after that the sites simply send all the remaining points to the coordinator.  


\begin{algorithm}[t]
\caption{Communication-Round Tradeoff under Horizontal Partition}
\label{alg:tradeoff}
\begin{algorithmic}[1]
\Require $S_i$ is initialized as the point set held by Site $i$; $r$  (\#communication-rounds) is a user-chosen parameter; $d$ is a parameter specified in the analysis.
\Ensure the global skyline

\State Site $i$ computes its local skyline points and prunes the other (dominated) points  in $S_i$

\State $\ell \leftarrow 0$; $t \gets \lceil r/2 \rceil$

\While{$(\ell \le t-1) \wedge (\exists i \in [s]\ s.t.\ S_i \neq \emptyset)$}

\State All sites and the coordinator jointly compute $(1/d, 1/2d)$-quantiles according to the $x$-coordinates of points in $\bigcup_{i \in [s]} S_i$ \label{ln:tradeoff-1}
\Comment the quantile points naturally partition the Euclidean plane to $d$ strips


\For{each $i\ s.t. \ S_{i} \neq \emptyset$}  \label{ln:tradeoff-1}
\State Site $i$, for each non-empty strip, sends the point with the largest $y$-coordinate to the coordinator
\EndFor 

\State The coordinator, for each strip, finds the point with the largest $y$-coordinate among points received from sites; let $Y$ denote the set of these points among all strips \label{ln:tradeoff-2}
 
\State The coordinator computes new skyline points from $Y$ and sends them to each site 

\For{each $i\ s.t. \ S_{i} \neq \emptyset$} 
\State Site $i$ prunes $S_i$ by new global skyline points received from the coordinator
\EndFor 

\State $\ell \leftarrow \ell + 1$

\EndWhile 

\State $\forall i \in [s]$, Site $i$ sends $S_i$ to the coordinator

\State The coordinator updates the global skyline using the new points received from sites
\end{algorithmic}
\end{algorithm}

We now show the correctness of Algorithm~\ref{alg:tradeoff} and analyze its costs.  The high level intuition on the round efficiency of Algorithm~\ref{alg:tradeoff} is that at each round, the point with the maximum $y$-coordinate in each strip will either contribute to the global skyline or help to prune all the points in that strip.  Compared with Algorithm~\ref{alg:optimal}, one can think that we are trying to prune the whole data set {\em in parallel}, that is, in each strip of the plane.  This will reduce the round complexity at the cost of mildly increasing the total communication cost.

\paragraph{Correctness} The correctness of Algorithm~\ref{alg:tradeoff} is straightforward: our skyline computation does not prune any point that is not dominated by others. Indeed, up to the $(t-1)$-th step (or, $(2t-2)$-th round), what Algorithm~\ref{alg:tradeoff} does can be summarized as ``sites send candidate global skyline points $\to$ the coordinator computes new global skyline points from these candidates $\to$ sites use new skyline points to prune their local datasets''. At the $(2t-1)$-th round, sites just send all the remaining unpruned points to the coordinator so that we will not miss any skyline points.  

\paragraph{Communication cost}   We count the communication cost in two parts.  The first part is the communication needed at the first $(t-1)$ steps, and the second part is the total number of remaining points at all sites after the $(t-1)$-th step, which will be sent to the coordinator all at once at the final round.

We first analyze the cost of computing quantiles at each step. We can compute $(\eps, \eps/2)$-quantiles using the following folklore algorithm: the $i$-th site (for all $i \in [s]$) sends the coordinator the exact $\eps/2$-quantiles $Q_i$ of its local point set $S_i$. Using $\{Q_1, \ldots, Q_s\}$, the coordinator can answer quantile queries as follows: Given a query rank $\beta$, it returns the largest $v$ satisfying 
$$ \textstyle \beta - \sum_{i \in [s]} rank_i(v) \ge 0,$$ 
where $rank_i(v) = n_i(v) \cdot (\eps/2 \cdot \abs{S_i})$, and $n_i(v)$ is the number of $\eps/2$-quantiles in $Q_i$ that is smaller than $v$. It is easy to see that $0 \le \beta - v \le \eps/2 \cdot \sum_{i \in [s]} \abs{S_i}$. The following lemma summarizes the communication cost of this algorithm.

\begin{lemma}
There is an algorithm that computes $(\eps, \eps/2)$-quantiles in the coordinator model using one round and $O(s/\eps)$ communication.
\end{lemma}

Thus the communication used for quantile computation can be bounded by $O(s d)$ at each step. The rest communication at each step includes sending local maximums at all strips and new skyline points, which can be bounded by $O(s d)$ we as well.  To sum up the total communication in the first part is bounded by $O(sd (t-1))$.

The rest of the analysis is devoted to the second part, that is, to bound the number of the remaining points after the $(t-1)$-th step.  

We first assume that the output size $k$ is known, and $d$ will be chosen as a function of $k$. We will then show how to remove this assumption.  

Let $Y_\ell \in [1,d]$ denote the number of new skyline points we find at the $\ell$-th step.  
Observe that in each strip, if the point with the largest $y$-coordinate is not a skyline point, then the rest of the points in that strip cannot be skyline points and thus are pruned.  After the first step, there are at most $Y_{1}$ strips having point and each strip has at most $2n/d$ points, so there are at most
\begin{equation*} \label{eq:cost1}
 Y_{1} \cdot 2n/d=2n Y_{1}/d
\end{equation*} 
points left.  After the second step, there are at most $Y_{2}$ strips having point and each strip has at most $2(2n Y_{1}/d)/d$ points, so there are at most 
\begin{equation*} \label{eq:cost2}
Y_{2} \cdot 2 \left( 2n Y_{1}/d \right) /d=4n Y_{1}Y_{2}/d^2 \qquad \left(Y_{1}+Y_{2} \le k \right)
\end{equation*} 
points left.  After the $(t-1)$-th step, there are at most 
\begin{eqnarray} 
&& 2^{t-1} n  \prod_{\ell \in [t-1]} Y_\ell \left/d^{t-1} \right. \qquad \left( \sum_{\ell \in [t-1]} Y_\ell \le k \right) \label{eq:cost3} \\
&\le& n \left( \frac{2k}{(t-1)d} \right)^{t-1} \label{eq:cost4}
\end{eqnarray}
points left, where from (\ref{eq:cost3}) to (\ref{eq:cost4}) we have used the AM-GM inequality and the equality holds when all $Y_\ell = {k}/{(t-1)} \ (\ell = 1, \ldots, t-1)$.
We thus have at most $n(2k/((t-1)d))^{t-1}$ points left at sites after $(t-1)$-th step, and the sites will send all of them in the final (i.e., $(2t-1)$-th) round.  Adding two parts together, the total communication cost is bounded by
\begin{equation}
\label{eq:cost5}
O(sd(t-1)) + n \left(\frac{2k}{(t-1)d}\right)^{t-1}.
\end{equation}
When 
\begin{equation}
\label{eq:cost6}
d = \frac{2k}{t-1} \cdot \left( \frac{n (t-1)}{2sk} \right)^{1/t},
\end{equation}
Expression (\ref{eq:cost5}) simplifies to be $O(sk^{(t-1)/t} (n/s)^{1/t})$.

\paragraph{Dealing with unknown $k$}  We now show how to deal with the case that we do not know $k$ at the beginning.  A simple idea is to guess $k$ as powers of $2$ (i.e., $1, 2, 4, 8, \ldots$), and for each guess, we run our algorithm, and report error if $\sum Y_\ell > k$ at some point, in which case we double the value of $k$ and rerun the algorithm.  The correctness of the algorithm still holds. The round complexity, however, may blow up by a factor of $\log k$ in the worst case.  We will show that there is a way to preserve the round complexity even when we do not know $k$ at the beginning.

The new idea is to guess $k$ progressively, based on the number of new skyline points found in the previous step.  More precisely, we set the guess of $k$ at the $\ell$-th step, denoted by $k_\ell$, to be $Y_{\ell-1} \cdot (t-1)$, for $\ell \ge 2$; and we set $k_1 = t - 1$ to begin with.  Now at the $\ell$-th step we use
\begin{equation}
\label{eq:guess-k-1}
d = d_\ell = \frac{2 k_\ell}{t-1} \cdot \left(\frac{n(t-1)}{2s}\right)^{1/t}
\end{equation}
strips for the pruning. Note that (\ref{eq:guess-k-1}) is very similar to (\ref{eq:cost6}), where we have replaced the first $k$ in (\ref{eq:cost6}) by $k_\ell$ and removed the second $k$ in (\ref{eq:cost6}).

Similar to (\ref{eq:cost4}), after $(t-1)$-th step, there are at most 
\begin{equation*}
\label{eq:guess-k-2}
 2^{t-1} n \prod_{\ell \in [t-1]} ({Y_\ell}/{d_\ell})
\end{equation*}
points left, and consequently the total communication cost is bounded by
\begin{equation}
\label{eq:guess-k-3}
s \sum_{\ell \in [t-1]}d_\ell + 2^{t-1} n \prod_{\ell \in [t-1]} ({Y_\ell}/{d_\ell}).
\end{equation}
We now bound the two terms in (\ref{eq:guess-k-3}) separately. We first have
\begin{eqnarray} 
s\sum_{\ell \in [t-1]} {d_\ell} &=& \frac{2s \sum_{\ell \in [t-1]} {k_\ell} }{(t-1)} \cdot \left(\frac{n(t-1)}{2s}\right)^{1/t} \nonumber \\ &\le& 2sk \cdot \left(\frac{n(t-1)}{2s}\right)^{1/t}.  \label{eq:guess-k-4}
\end{eqnarray} 
where we have used the inequality
\begin{equation*}
\sum_{\ell \in [t-1]} k_\ell = (t-1)\left(1+\sum_{\ell \in [t-2]}{Y_\ell}\right) \le (t-1)k.
\end{equation*}
For the second term, we have 
\begin{eqnarray}
2^{t-1} n \prod_{\ell \in [t-1]} (Y_\ell /d_\ell) &=& n \frac{Y_{1}Y_{2} \cdots Y_{t-1}}{Y_{1}Y_{2} \cdots Y_{t-2}} \left(\frac{n(t-1)}{2s}\right)^{t/(t-1)} \nonumber \\
&=& n Y_{t-1} \cdot \frac{2s}{n(t-1)}  \left(\frac{n(t-1)}{2s}\right)^{1/t} \nonumber \\
&\le& \frac{2sk}{(t-1)} \left(\frac{n(t-1)}{2s}\right)^{1/t} \label{eq:guess-k-5}.
\end{eqnarray}
By (\ref{eq:guess-k-4}) and (\ref{eq:guess-k-5}), the total cost is bounded by $O(sk (n/s)^{1/t}) = O(sk (n/s)^{1/\lceil r/2 \rceil})$, as claimed.

\paragraph{Running time}
The time cost at each site involves three parts: the computation of the local skyline, the computation of local quantiles, and the point prunings. Computing local skyline again cost $O(n_i \log n_i)$ where $n_i$ is the number of points at the $i$-th site.  The cost of point prunings can again be made linear in $n_i$ for the same reason as that in Algorithm~\ref{alg:optimal}.  Now we analysis the time cost of computing the local quantiles.  Since points are sorted after the local skyline computation, computing (exact) local quantiles needs $O(d_\ell)$ time at the $\ell$-th step.  Thus the total time is bounded by $\sum_{\ell \in [t-1]} d_\ell = O(k (n/s)^{1/t}) = O(k (n/s)^{1/\lceil r/2 \rceil})$. 

The running time at the coordinator also consists of three parts: the computation of approximate global quantiles at each step,  the computation of new skyline points from the first step to the $(t-1)$-th step, and the computation of skyline points (output) at the end.  The observation is that at each step, for each of the three tasks, the running at the coordinator can be asymptotically bounded by the number of points it receives from all sites in that step, and thus the total running time at the coordinator is asymptotically upper bounded by the total communication cost.

\subsubsection{An Algorithm for Sorted Datasets}
\begin{figure}[t]
\centering
\includegraphics[height = 1.2in]{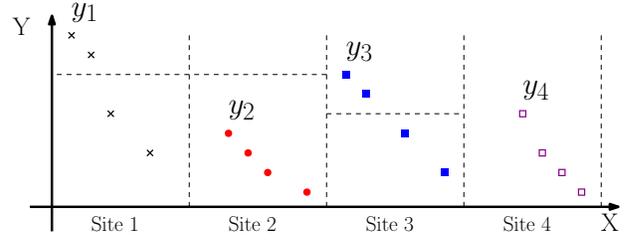}
\caption{Algorithm for Sorted Dataset}
\label{fig:sorted}
\end{figure}

In this section we show that we can do much better if the data points are partitioned to the $s$ sites in a {\em sorted} order with respect to the $x$-coordinate (or, the $y$-coordinate). In other words, let $x_1 \le x_2 \le \ldots \le x_{s-1}$ be $(s - 1)$ split points, and let $x_0 = -\infty$ and $x_s = \infty$.  The $i$-th site gets all points between $(x_{i-1}, x_i]$.  See Figure~\ref{fig:sorted} for an illustration.   As mentioned in the introduction, in the coordinator model we cannot afford to repartition the dataset. Therefore the algorithm presented in this section is only useful for settings where the data has already been sorted among the sites.

\begin{theorem}
\label{thm:sorted}

For $n$ points in the $2$-dimensional Euclidean space partitioned among the $s$ sites in the coordinator model in the sorted manner according to their $x$-coordinates or $y$-coordinates, there exists an algorithm for computing the skyline that uses $O(k+s)$  communication and $2$ rounds, where $k$ is the output size, that is, the number of points in the skyline.  The total running time at the $i$-th site is $O(n_i \log n_i)$ where $n_i$ is the number of points at the $i$-th site, and that at the coordinator is $O(k+s)$.
\end{theorem}

\begin{algorithm}[t]
\caption{Sorted Datasets under Horizontal Partition}
\label{alg:sorted}
\begin{algorithmic}[1]
\Require $S_i$ is initialized as the point set held by Site $i$
\Ensure the global skyline

\State Site $i$ computes its local skyline points and discards the other points in $S_i$

\State Site $i$ sends the point with the largest $y$-coordinate, denoted by $y_i$, to the coordinator 

\State The coordinator computes for $\forall i \in [s], z_i = \max \{y_{i+1}, \ldots, y_s \}$, and sends $z_i$ to Site $i$
 
\State Site $i$ prunes $S_i$ using $z_i$, and sends the rest points to the coordinator

\State The coordinator updates the global skyline using the new points received from sites
\end{algorithmic}
\end{algorithm}

We described our algorithm for sorted datasets in Algorithm~\ref{alg:sorted}.  Let us explain in words. Same as before, each site first does a local pruning and computes its local skyline. In the first round, the $i$-th site sends the coordinator the point which has the largest $y$-coordinate, denoted by $y_{i}$. In the second round, the coordinator for each $i \in [s]$ computes the value $z_i$ which is the maximum value among $\{y_{i+1}, \ldots, y_{s}\}$, and sends $z_{i}$ to the $i$-th site. The $i$-th site then prunes all its local points with  $y$-coordinate smaller than $z_i$, and sends the rest of its points to the coordinator. 


We now show the correctness of this algorithm and analyze its costs. The claim is that the points in $S_i$ with $y$-coordinate larger than $z_{i}$, denoted by $P_i$, must on the global skyline. Indeed, points in $P_i$ can not be dominated by any point in $S_{1}, \ldots, S_{i-1}$ since all points in $S_{i}$ have  $x$-coordinates larger than those points in $S_{1}, \ldots, S_{i-1}$. On the other hand, points in $P_i$ cannot be dominated by any point in $S_{i+1},...,S_{s}$ since all points in $P_i$ have $y$-coordinates larger than the those points in $S_{i+1}, \ldots, S_{s}$.  
The communication of the algorithm includes sending $y_{i}$ and $z_{i}$ (costs $2s$) plus sending skyline points (costs $k$).   The running time at each site is dominated by the local skyline computation, and the time cost at the coordinator is clearly $O(k+s)$ ($O(s)$ for the first round and $O(k)$ for the second round).

\subsection{Lower Bounds}
\label{sec:lb}

\subsubsection{Infinite Rounds}
\label{sec:lb-infinite}

We prove a lower bound for the infinite-round case by a reduction from a communication problem called \DISJ.  The lower bound matches the upper bound by Algorithm~\ref{alg:optimal} up to a logarithmic factor which counts the number of bits used to represent a point in the Euclidean plane.

In \DISJ, each of the $s$ sites gets an $m$-bit vector. Let $X_i = (X_i^1, \ldots, X_i^m)$ be the vector the the $i$-th site gets.   We can view the whole input as an $s \times m$ matrix $X$ with $X_i\ (i \in [s])$ as rows.  The \DISJ\ problem is defined as follows:
\begin{eqnarray*}
\begin{array}{l}
\text{\DISJ}(X_1, \ldots, X_s)  =  \left\{
  \begin{array}{rl}
   1, & \text{if there exists a $j \in [m]$  s.t.} \\ 
    & \forall i \in [s], X_i^j = 1,\\
   0, & \text{otherwise.}
  \end{array}
  \right.
\end{array}
\end{eqnarray*} 

\begin{lemma}[\cite{BEOPV13}]
Any randomized algorithm for \DISJ\ that succeeds with probability $0.51$ has communication cost $\Omega(sm)$. The lower bound holds even when we allow an infinite number of communication rounds.
\end{lemma}

\begin{figure}[t]
\centering
\includegraphics[height = 1.7in]{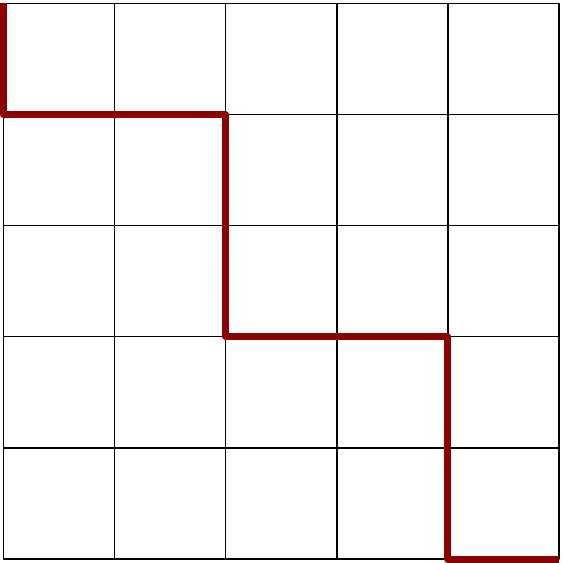}
\caption{Translating vector $X'_1 = 1001100110\ (m = 5)$ to a staircase.}
\label{fig:reduction}
\end{figure}

\paragraph{The Reduction}
Given the $m$-bit vector $X_i$ for \DISJ, the $i$-th site first converts it to a $2m$-bit vector $X'_i$ as follows: each $0$ bit will be converted to $01$, and each $1$ bit will be converted to $10$.  For example, when $m = 5$ and $X_i = 10101$, $X'_i$ should be $1001100110$.  The next step is to convert $X'_i$ to a staircase.  This step is illustrated in Figure~\ref{fig:reduction}.  We can ``embed'' the staircase into an $m \times m$ grid.  The staircase starts from the top-left point of the grid, and grows in $2m$ steps.  In the $\ell$-th step, if the $\ell$-th coordinate of $X'_i$ is $0$, then the staircase grows one step horizontally rightwards; otherwise if the $\ell$-th coordinate is $1$, then the staircase grows one step vertically downwards.

\begin{figure}[t]
\centering
\includegraphics[height = 1.7in]{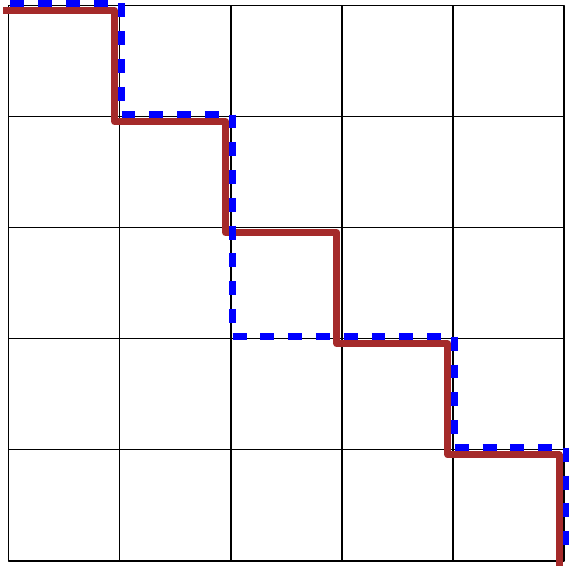}
\caption{The solid red skyline corresponds to the case that \DISJ$(X_1, \ldots, X_s) = 0$, and the dash blue skyline corresponds to the case that \DISJ$(X_1, \ldots, X_s) = 1$.}
\label{fig:skylines}
\end{figure}

The observation is that if we create $s$ staircases using $X'_1, \ldots, X'_s$, then the skyline of the union of these $s$ staircases is closely related to the value of \DISJ$(X_1, \ldots, X_s)$: If \DISJ$(X_1, \ldots, X_s) = 0$, then the skyline will be in the form of the red curve in Figure~\ref{fig:skylines}; otherwise, the skyline will be different from the red curve (e.g., be the blue curve in Figure~\ref{fig:skylines} if the $3$rd coordinates of $X_1, \ldots, X_s$ are all $1$).  This is because for each column $j \in [m]$ in the grid, as long as there is one $i \in [s]$ such that the $j$-th coordinate of $X_i$ is $0$, or the $(2j-1)$-th and $(2j)$-th coordinates of $X'_i$ is $01$, the skyline within the $j$-th column of the grid will be like ``$\urcorner$''; otherwise if for all $i \in [s]$ the $j$-th coordinate of $X_i$ is $0$, then the skyline within the $j$-th column of the grid will be like ``$\llcorner$''.
The other direction also holds, that is, if the skyline is in the form of the red curve, then \DISJ$(X_1, \ldots, X_s) = 0$; otherwise, \DISJ$(X_1, \ldots, X_s) = 1$.
\smallskip

When the size of the skyline is $k$, we set $m = k$ according to our reduction, and obtain the following theorem.
\begin{theorem}
\label{thm:lb-infinite}

Any randomized protocol for computing skyline in the coordinator model with horizontal partition that succeeds with probability $0.51$ has communication cost $\Omega(s k)$ bits, where $k$ is number of points in the skyline.   The lower bound holds even when we allow an infinite number of communication rounds.
\end{theorem}

\subsubsection{One Round}
\label{sec:lb-one}

We now prove an $\Omega(n)$ communication lower bound for the case that the algorithm needs to finish in one round.  This proof is simpler than the infinite-round case -- we only need two sites to participate in the game. We assign Site $1$ an $m$-bit vector $u$, which can be converted to a staircase in the $m \times m$ grid just like the infinite-round case, and assign Site $2$ one bit $v$, which is translated to the upper-right conner point of the grid if $v = 1$ and the lower-left conner point of the grid if $v = 0$.  We have the following simple observation, which holds because if we change one bit of the vector $m$ from $0$ to $1$, we will change a ``$\urcorner$'' to ``$\llcorner$'' in the staircase, and thus change the skyline; same for replacing a bit $1$ to $0$.

\begin{observation}
\label{ob:lb}

If the coordinator needs to compute the global skyline, and $v = 0$, then it needs to learn the vector $u$ exactly.
\end{observation}

We immediately have the following lemma.
\begin{lemma}
Any one round randomized algorithm for the coordinator to learn $u$ exactly with probability $0.51$ has communication cost $\Omega(m)$.
\end{lemma}

Note that when $v = 1$, the skyline only consists of a single point in the upper right conner. Therefore the output size $k$ is not directly related to the value $m$; we thus can set $m = \Omega(n)$.

\begin{theorem}
\label{thm:lb-one}

Any one round randomized algorithm for computing skyline in the coordinator model with horizontal partition that succeeds with probability $0.51$ has communication cost $\Omega(n)$ bits, where $n$ is the total number of points held by all sites.
\end{theorem}

\section{Vertical Partition}
\label{sec:vertical}

In the vertical partition, each site holds a single coordinate of all the data points.   Since we consider points in the $2$-dimensional Euclidean space in this paper, we only have two sites, named Alice and Bob.  We store the ID (key) of each point on both sites. More precisely, Alice has a set of tuples $\{(x_{p_1}, \ID_{p_1}),  (x_{p_2}, \ID_{p_2}), \ldots \}$, where $x_{p_i}$ is the $x$-coordinate of point $p_i$ and $\ID_{p_i}$ is the ID of $p_i$; and Bob has a set of tuples $\{(\ID_{q_1}, y_{q_1}),  (\ID_{q_2}, y_{q_2}), \ldots \}$, where $y_{q_i}$ is the $y$-coordinate of point $q_i$.  We can join the two tables to recover the information of all points.  

\subsection{A Strong Lower Bound}
\label{sec:lb-vertical}

As mentioned in the introduction, in the vertical partition case there is a strong communication lower bound, which states that in the worst case sites have to send almost everything to the coordinator.
We prove the lower bound by a reduction from a classical problem in communication complexity called two-party set-disjointness (denoted by \TDISJ). In the \TDISJ, we have two parties Alice and Bob.  Alice holds a set of numbers $A \subseteq [n]$, and Bob holds a set of numbers $B \subseteq [n]$, and they want to jointly compute  
\begin{eqnarray*}
\begin{array}{l}
\text{\TDISJ}(A, B)  =  \left\{
  \begin{array}{rl}
   1, & A \cap B \neq \emptyset, \\
   0, & \text{otherwise.}
  \end{array}
  \right.
\end{array}
\end{eqnarray*} 

\begin{fact}[\cite{BJKS04}]
\label{thm:2DISJ}
Any randomized protocol for \TDISJ\ that succeeds with probability $0.51$ has communication cost $\Omega(n)$.
\end{fact}

\paragraph{The Reduction}
Alice uses her input $A$ to create the $x$-coordinates of the $n$ points as follows: for each $a \in A$, she creates a point $(2, a)$ where $2$ is the $x$-coordinate and $a$ is the ID of the point; and for each $a \in [n] \backslash A$, she creates a point $(1, a)$.  Bob uses his input $B$ to create the $y$-coordinates of the $n$ points in a similar way:  for each $b \in B$, he creates a point $(b, 2)$ where $2$ is the $y$-coordinate and $b$ is the ID of the point; and for each $b \in [n] \backslash B$, he creates a point $(b, 1)$. 
\smallskip

It is easy to see that if $A \cap B \neq \emptyset$, then there is at least one point in the $(2, 2)$ position, which will be the only point in the skyline; otherwise, the skyline will consist of two points $(1, 2)$ and $(2, 1)$.  By Fact~\ref{thm:2DISJ} we have

\begin{theorem}
\label{thm:lb-vertical}
Any randomized algorithm for computing skyline in the distributed model with vertical partition that succeeds with probability $0.51$ has communication cost $\Omega(n)$, regardless how many rounds the algorithm uses.
\end{theorem}

\subsection{A Simple Heuristic}
\label{sec:heuristic}

Given the strong lower bound result in Section~\ref{sec:lb-vertical}, we propose in this section a heuristic which is  communication efficient on real-world datasets.  Our algorithm can be thought as a batched pruning version of the Threshold-Algorithm based approach \cite{BGZ04,LYLC06}, but has been tailored for bounded-round distributed computation.  We call our heuristic {\em interactive pruning}; the pseudocode is described in Algorithm~\ref{alg:vertical}.

\begin{algorithm}[t]
\caption{Interactive Pruning under Vertical Partition}
\label{alg:vertical}
\begin{algorithmic}[1]
\Require $S_1$ is the set of the $x$-coordinates of points held by Alice, and $S_2$ is the set of the $y$-coordinates of points held by Bob.  $\rho$ (number of groups) and $r$ (round budget) are two user-chosen parameters.
\Ensure the global skyline
\smallskip
\State Alice sorts $S_1$ decreasingly and then partitions it to $\rho$ groups of equal size, denoted by $\G_x = \{G_{x1}, \ldots, G_{x \rho}\}$.  Bob does the same thing on $S_2$ and gets groups $\G_y = \{G_{y1}, \ldots, G_{y\rho}\}$.  Alice and Bob send the splitting coordinates of the group partitions to the coordinator.
Given a point $p$, let $g_x(p)$ be the index of the group in $\G_x$ that contains $p$, and $g_y(p)$ be the index of the group in $\G_y$ that contains $p$

\State The coordinator maintains a set of points $R$ denoting the recovered points, and four variables: $f_x, f_y, l_x, l_y$.

\State The coordinator communicates with Alice and Bob to learn both the $x$-coordinates and $y$-coordinates of points in $G_{y1}$ and $G_{x1}$, adds them to $R$, and then updates $l_x \gets \min\{g_x(p) \ |\ p \in G_{y1}\} + 1$ and $l_y \gets \min\{g_y(p) \ |\ p \in G_{x1}\} + 1$  \label{ln:vertical-1}

\State The coordinator sets $f_{x} \gets 2$, $f_{y} \gets 2$ 

\State $ j \leftarrow 1$ \Comment index of steps; one step uses two rounds    \label{ln:vertical-2}

\While{$(2j \le r - 4) \wedge (l_x > f_x)  \wedge (l_y > f_y)$} \label{ln:vertical-3}

\State If $l_x - f_x \le l_y - f_y$, the coordinator requests group $G_{xf_x}$ from Alice and updates $f_x \gets f_x + 1$, else it requests $G_{yf_y}$ from Bob and updates $f_y \gets f_y + 1$

\State The coordinator recovers and adds points in $G_{xf_x}$ (or  $G_{yf_y}$) to $R$, and updates $l_y \gets \min\{l_y, \min\{g_y(p) \ |\ p \in G_{xf_x}\} + 1\}$ (or $l_x \gets \min\{l_x, \min\{g_x(p) \ |\ p \in G_{yf_y}\} + 1\}$)
 
\State $j \gets j + 1$
\EndWhile  \label{ln:vertical-4}

\If{$(l_x > f_x )  \wedge (l_y > f_y)$}
\State The coordinator compares $l_x - f_x$ and $l_y - f_y$; w.l.o.g., assume $l_x - f_x$ is smaller. The coordinator sends $f_x$ and $l_x$ to Alice. Alice then sends the $x$-coordinates and IDs of points (denoted by $P$) in $G_{xf_x}, \ldots, G_{x (l_x-1)}$ to the coordinator.  
\State The coordinator sends the IDs of points in $P$ to Bob. Bob then sends the $y$-coordinates of points in $P$ to the coordinator. The coordinator adds points in $P$ to $R$
\EndIf

\State The coordinator computes and outputs the skyline of $R$
\end{algorithmic}
\end{algorithm}

We now explain our algorithm in words. Let $r$ be the round budget and $\rho$ be the parameter denoting the number of groups we use for the algorithm.  We say a point $p$ is {\em recovered} at the coordinator if both the $x$ and $y$ coordinates of $p$ are known to the coordinator. 

Alice and Bob first sort decreasingly all points according to the $x$-coordinates and $y$-coordinates respectively, and then partition the points to $\rho$ groups, denoted by $G_{x1},  \ldots, G_{x\rho}$ and $G_{y1}, \ldots, G_{y\rho}$ respectively.  The coordinator tries to learn both coordinates of the points that possibly lie in the skyline progressively.  It maintains four variables $f_x, f_y, l_x, l_y$: $f_x$ and $f_y$ stand for the indices of the next groups Alice and Bob will send to the coordinator respectively; and groups with indices at least $l_x$ and $l_y$ at Alice and Bob respectively have already been pruned.  Observe that if $f_x \ge l_x$ or $f_y \ge l_y$, then the coordinator must have recovered all the skyline points, and thus the algorithm can terminate.  Based on this fact, the coordinator always chooses the site with the smaller {\em gap value} $l_x - f_x$ or $l_y - f_y$, and asks for the information of its next group $G_{xf_x}$ or $G_{yf_y}$.

Algorithm~\ref{alg:vertical} proceeds in three stages.  The first stage consists of two rounds (Line \ref{ln:vertical-1} to \ref{ln:vertical-2}). The goal of the first stage is to recover the first groups $G_{x1}$ and $G_{y1}$ at Alice and Bob, and update the variables $l_x$ and $l_y$ to determine the next group to recover.  The second stage consists of $\lfloor (r - 4)/2 \rfloor$ steps each of which consisting of two rounds (Line \ref{ln:vertical-3} to \ref{ln:vertical-4}). The goal of the second stage is to recover points progressively from the site which has a smaller gap value.   The algorithm will stop early if at some step all points have been either pruned or recovered at a site. The goal of the third stage is to recover all groups that are still {\em not} recovered or pruned after the first two stages.  This is done at the coordinator by first asking the site with the smaller gap value to get the IDs of points in these groups together with one of the two coordinates, and then contacting the other site for the other coordinates of these points.  We include a running example for Algorithm~\ref{alg:vertical} in Appendix~\ref{sec:example}.

The correctness of the algorithm is straightforward since when our algorithm stops, all points at one site are either recovered or pruned.  The local running time at each site is dominated by the initial sorting which is $O(n \log n)$, and the time cost at the coordinator is dominated by the final skyline computation which is also $O(n \log n)$. 

We comment that $\rho$ is a parameter that we need to choose at the beginning of the algorithm; we will discuss how to choose $\rho$ in the experiments in Section~\ref{sec:exp}.

\paragraph{A comparison to previous algorithms}
Our algorithm, the \bds\ algorithm and its improvement the \ids\ algorithm proposed in~\cite{BGZ04}, and the \pds\ algorithm proposed~\cite{LYLC06} are all based on the Threshold-Algorithm (TA)~\cite{FLN03}.  A clear difference is that \bds, \ids\ and \pds\ recover one point in each step at the coordinator, while we recover at least one group of points in each step which helps to significantly reduce the round cost.  The second major difference is that to improve the basic TA-base algorithm \bds, both \ids\ and \pds\ share the idea of estimating the {\em most probable terminating point}, and then using it to decide which site to access next.  \ids\ finds the most probable terminating point by calculating the score of each point which is the sum of the differences between each of its coordinates and the coordinate of the last point recovered by the sorted access at the respective site.   \pds\ uses linear regression to estimate the rank of each point, and the point with the lowest rank is considered to be the most probable terminating point.  
In our algorithm, instead of looking for the most probable terminating point, we focus on the number of unrecovered and unpruned groups remaining at each site.  We choose to request a new group of points from the site with the smaller number of remaining groups, with the purpose of terminating the algorithm earlier.  

It is well-known that the Threshold-Algorithm~\cite{FLN03} is {\em instance optimal} in terms of number of points probed, and thus if we only measure the number of points recovered at the coordinator in the process, our batched pruning approach has no advantage compared with the individual point prunings in \bds, \ids\ and \pds. However, as we shall see in the experiments (Section~\ref{sec:exp}), batched pruning can not only significantly reduce the round cost, but also bring down the overall communication cost in some cases.  This is because in each step the coordinator needs to send a message to request points from the sites in the sorted access, and consequently the round cost will  influence the communication cost as well.

\section{Experiments}
\label{sec:exp}

In this section we present the experimental studies of our proposed algorithms.  We have implemented all algorithms in C++. All experiments were conducted on a laptop with Inter Core i7 running Windows 7 with 4GB memory.

\paragraph{The Datasets}  We use both synthetic and real-world datasets.  We generated three synthetic datasets following the standard literature \cite{BKS01}.  The data partition among the sites will be described in Section~\ref{sec:exp-hor} (horizontal partition) and Section~\ref{sec:exp-ver} (vertical partition) respectively.  
\begin{itemize}
\item \ind (independent): We generate 20 million points. For each point we generate each of its coordinates independently uniformly at random from $[0,1]$

\item \cor (correlated): We select 200 thousand lines (denoted by $L_1$) perpendicular to the line from $(0,0)$ to $(1,1)$, where the intersections follow the standard normal distribution. Next, for each line in $L_1$ we pick $100$ points also following the standard normal distribution.  We have 20 million points in total.

\item \anti (anti-correlated):  We select 200 thousand lines (denoted by $L_2$) perpendicular to the line from $(0,1)$ to $(1,0)$, where the intersections follow the standard normal distribution. Next, for each line in $L_2$ we pick $100$ points also following the standard normal distribution.  We have 20 million points in total.
\end{itemize}
We make use of the following real-world datasets.
\begin{itemize}
\item \air: this dataset contains 3 million the airline itineraries and prices between 30 U.S. major cities in 2015 first quarter.\footnote{Available at \url{http://www.transtats.bts.gov}.}  We choose (minus) {\em fare} and (minus) {\em fare-per-mile} as the two attributes in $x$ and $y$ coordinates.  The skyline points are considered to be economic flights. 

\item \house~\cite{Lichman13}: this dataset contains 2 million household electric power consumption records gathered between December 2006 and November 2010.  We choose {\em voltage} and {\em intensity} as the two attributes in $x$ and $y$ coordinates.  The skyline points represent those households that are recommended to pay attention to the energy efficiency.

\item \cover~\cite{Lichman13}: this dataset contains 500  thousand natural statistics from four wilderness areas located in the Roosevelt National Forest of northern Colorado.  We choose {\em elevation} and {\em slope} as the two attributes in $x$ and $y$ coordinates. The skyline points represent areas that may have interesting geological behaviors.
\end{itemize}

\begin{figure*}[t]
\begin{minipage}[d]{0.33\linewidth}
\centering
\includegraphics[width=1\textwidth]{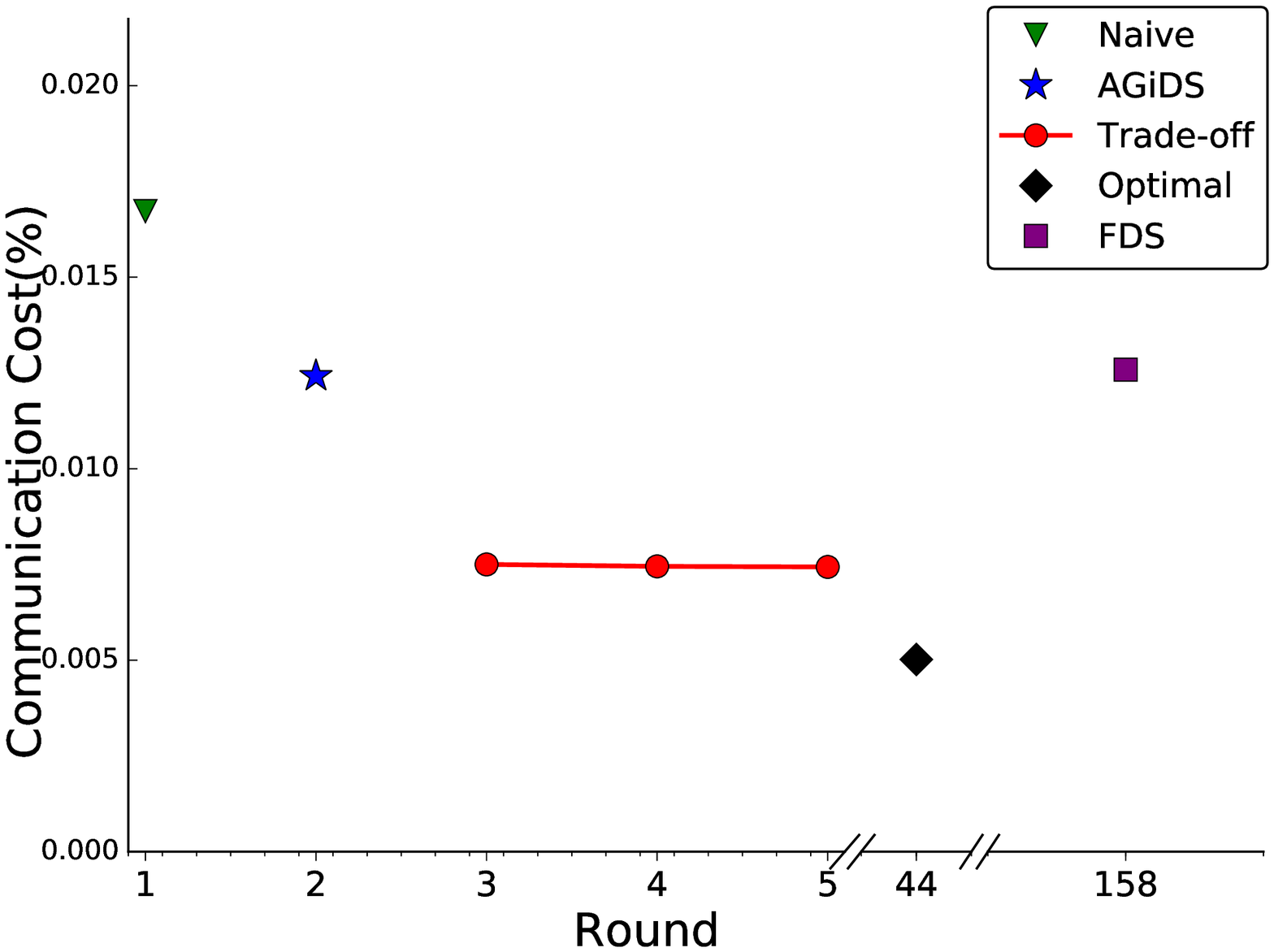}
\centerline{\anti}
\end{minipage}
\begin{minipage}[d]{0.33\linewidth}
\centering
\includegraphics[width=1\textwidth]{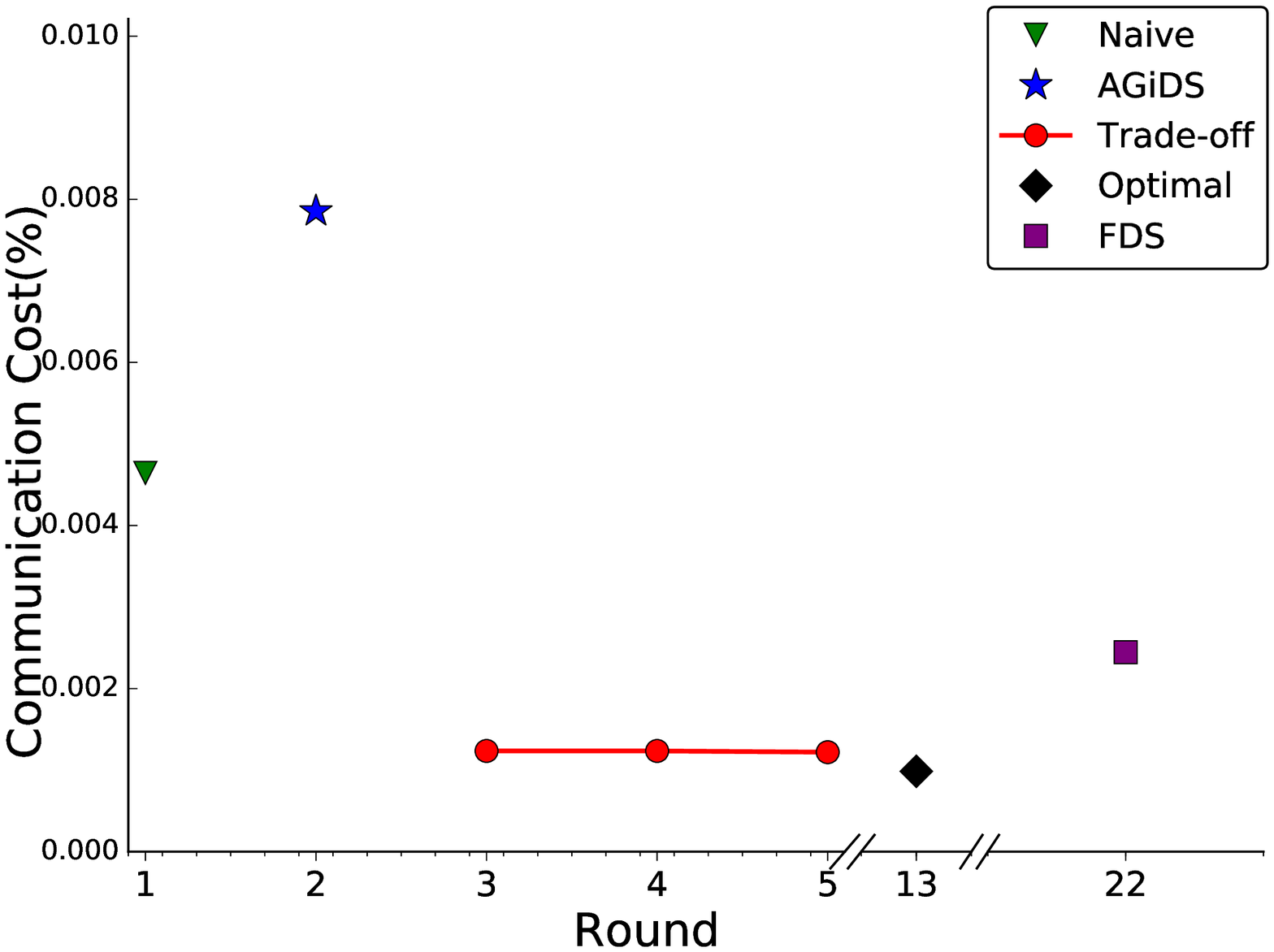}
\centerline{\ind}
\end{minipage}
\begin{minipage}[d]{0.33\linewidth}
\centering
\includegraphics[width=1\textwidth]{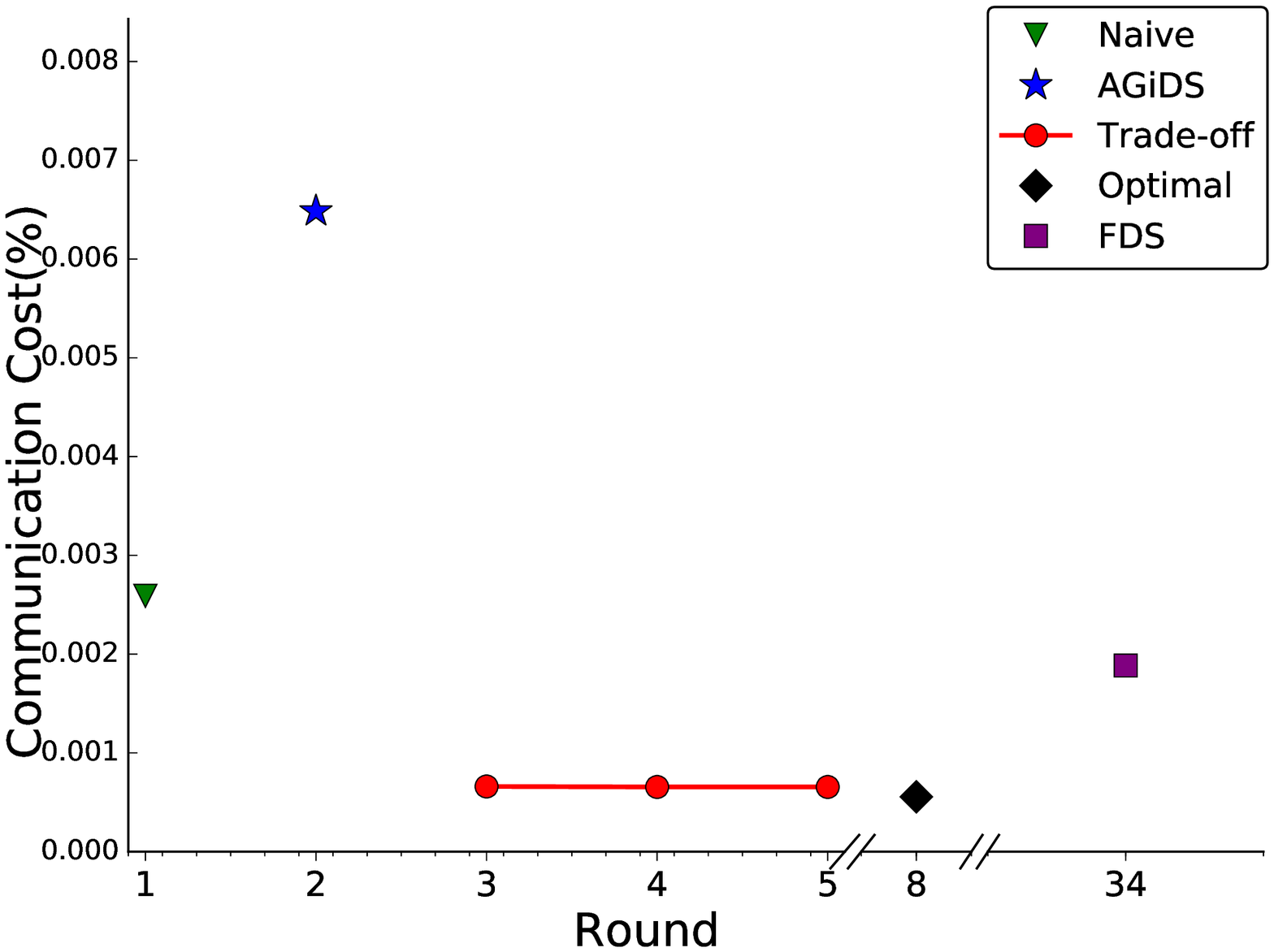}
\centerline{\cor}
\end{minipage}
\caption{Communication and round costs on synthetic datasets under horizontal partition}
\label{fig:horrsyn}
\end{figure*}

\begin{figure*}[t]
\begin{minipage}[d]{0.33\linewidth}
\centering
\includegraphics[width=1\textwidth]{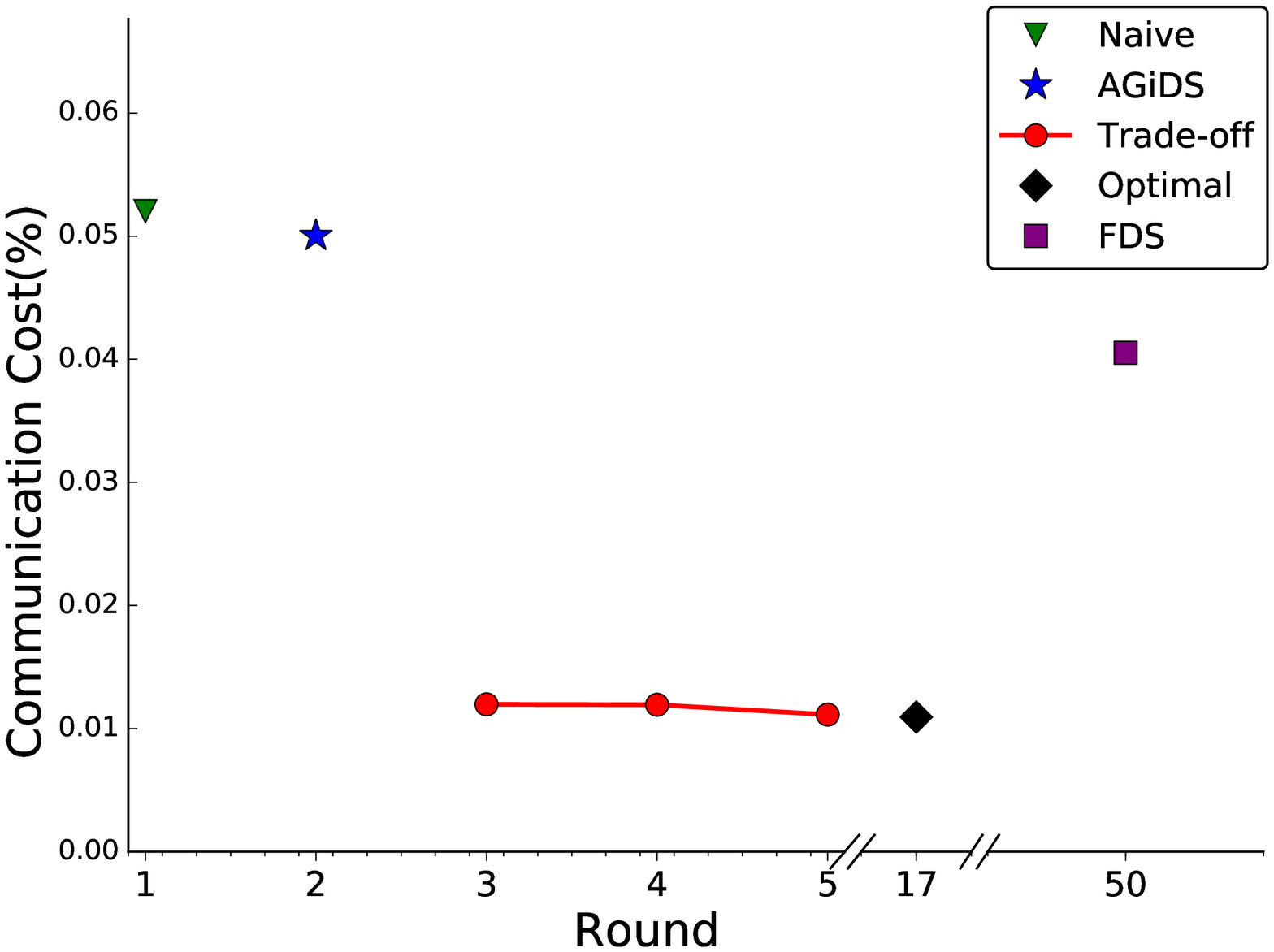}
\centerline{\air}
\end{minipage}
\begin{minipage}[d]{0.33\linewidth}
\centering
\includegraphics[width=1\textwidth]{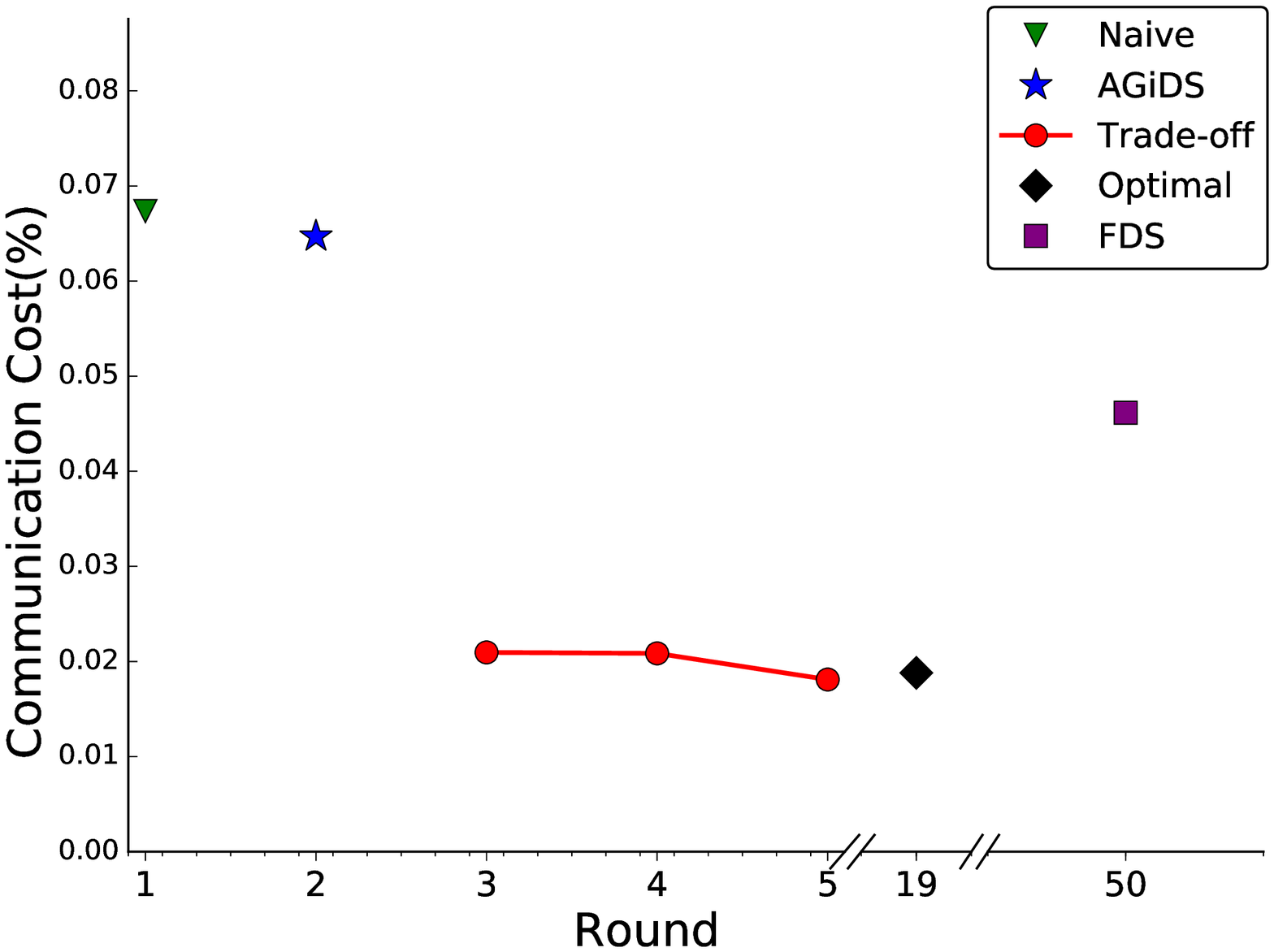}
\centerline{\house}
\end{minipage}
\begin{minipage}[d]{0.33\linewidth}
\centering
\includegraphics[width=1\textwidth]{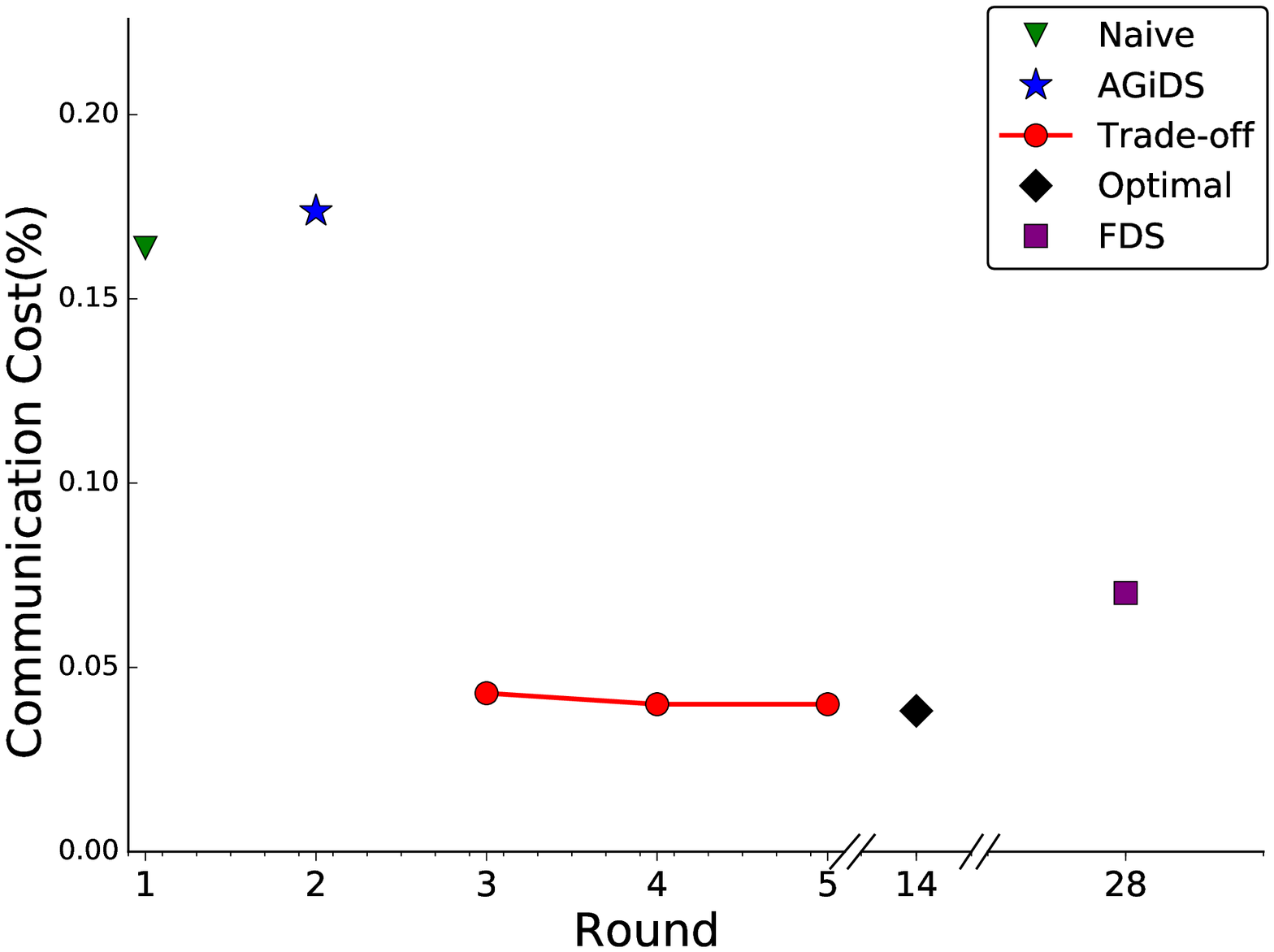}
\centerline{\cover}
\end{minipage}
\caption{Communication and round costs on real-world datasets under horizontal partition}
\label{fig:horrreal}
\end{figure*}

\subsection{Horizontal Partition}
\label{sec:exp-hor}

\paragraph{Algorithms} We compare the following algorithms in the case of horizontal partition.
\begin{itemize}
\item \naive:  the single round algorithm in which each site computes and sends its local skyline to the coordinator for a merge.

\item \opt: Algorithm~\ref{alg:optimal} in Section~\ref{sec:optimal}.  The algorithm that achieves the optimal communication cost.

\item \tradeoff: Algorithm~\ref{alg:tradeoff} in Section~\ref{sec:tradeoff}.  The algorithm that gives a smooth communication-round tradeoff.

\item \agids: We use the \agids\  algorithm proposed in \cite{RVDN09} as a comparison.   To make \agids\ fit in our model, we use the following version of the original algorithm: At the beginning, the coordinator and sites share the information of a grid in which each cell represents a range in the $x$/$y$-axes (equal width partition).  In the first round (the {\em planing phase}), sites send the information of non-empty cells to the coordinator.  In the second round (the {\em execution phase}), the coordinator finds the cells that may contribute to global skyline and sends the information to sites, and then sites send points in these cells to the coordinator.  Finally the coordinator computes the skyline of received points as the output.  We comment that this algorithm is also very similar to the {\em relaxed skyline} algorithm proposed in \cite{AKSU15}. We choose the grid to be $20 \times 20$ (total 400 cells) in our experiments which works the best under our settings.  

\item \fds: We use the \fds\  algorithm proposed in \cite{ZTZ09} as a comparison. The original algorithm proceeds in {\em iterations}. To make it fit in our model, we use three rounds for each iteration: In the first round (the {\em voluntary phase}), each site sends the top $\kappa$ points with the largest scores ($x+y$) to the coordinator. In the second round (the {\em compulsory and computation phase}),  the coordinator calculates the minimum score (denoted by $F_{\min}$) from received points and sends it to each site. Each site then sends all its local points that have  larger scores than $F_{\min}$ to the coordinator. The coordinator updates the global skyline with points received in the first two rounds. In the third round (the {\em feedback phase}), the coordinator calculates and sends each site a feedback, which consists of points that are guaranteed to dominate at least $\ell$ points in that site. And then each site does a local pruning. $\kappa$ and $\ell$ are two parameters in \fds; we choose the optimal values $\kappa=1$ and $\ell=1$ as reported in \cite{ZTZ09}. 
\end{itemize}

\paragraph{Data Partition}
For the three synthetic datasets, we partition points randomly to $20$ sites.  For \air, we partition the data records in the same city to the same site; we thus have $30$ sites.  For \house, we partition the data collected in every two consecutive months to the same site; $20$ sites in total.  For \cover, we just randomly partition the data records to $20$ sites.

\paragraph{Results and Discussions}
Figure~\ref{fig:horrsyn} and Figure~\ref{fig:horrreal} show the communication and round costs of the five tested algorithms on the three synthetic datasets and three real-world datasets mentioned above.  

We observed that by using three rounds, the communication cost of \tradeoff\ is $25\%$-$44\%$ of that of \naive\ on synthetic datasets and $22\%$-$31\%$ on real-world datasets, and is also close to \opt\ which uses more rounds.   We noticed that the advantage of \tradeoff\ against \naive\ is larger in real-world datasets (even of smaller sizes) than synthetic ones. This is because it is more likely in real-world datasets that a point in one site dominates most of the local skyline points in another site.  We observe that in \house, by using $5$ rounds \tradeoff\ needs even less communication than \opt. This is possible since \opt\ is just asymptotically optimal in the {\em worst case}.

The communication cost of \agids, which uses two rounds, is similar to \naive\ on \air\ and \house\ and even worse on \ind, \cor, and \cover, and is consequently much worse than \tradeoff\ and \opt.  \fds\ uses the largest amount of rounds (even more than \opt), but its communication is worse than both \tradeoff\ and \opt\ on all synthetic and real-world datasets.  

\begin{figure}[t]
\centering
\includegraphics[height = 2in]{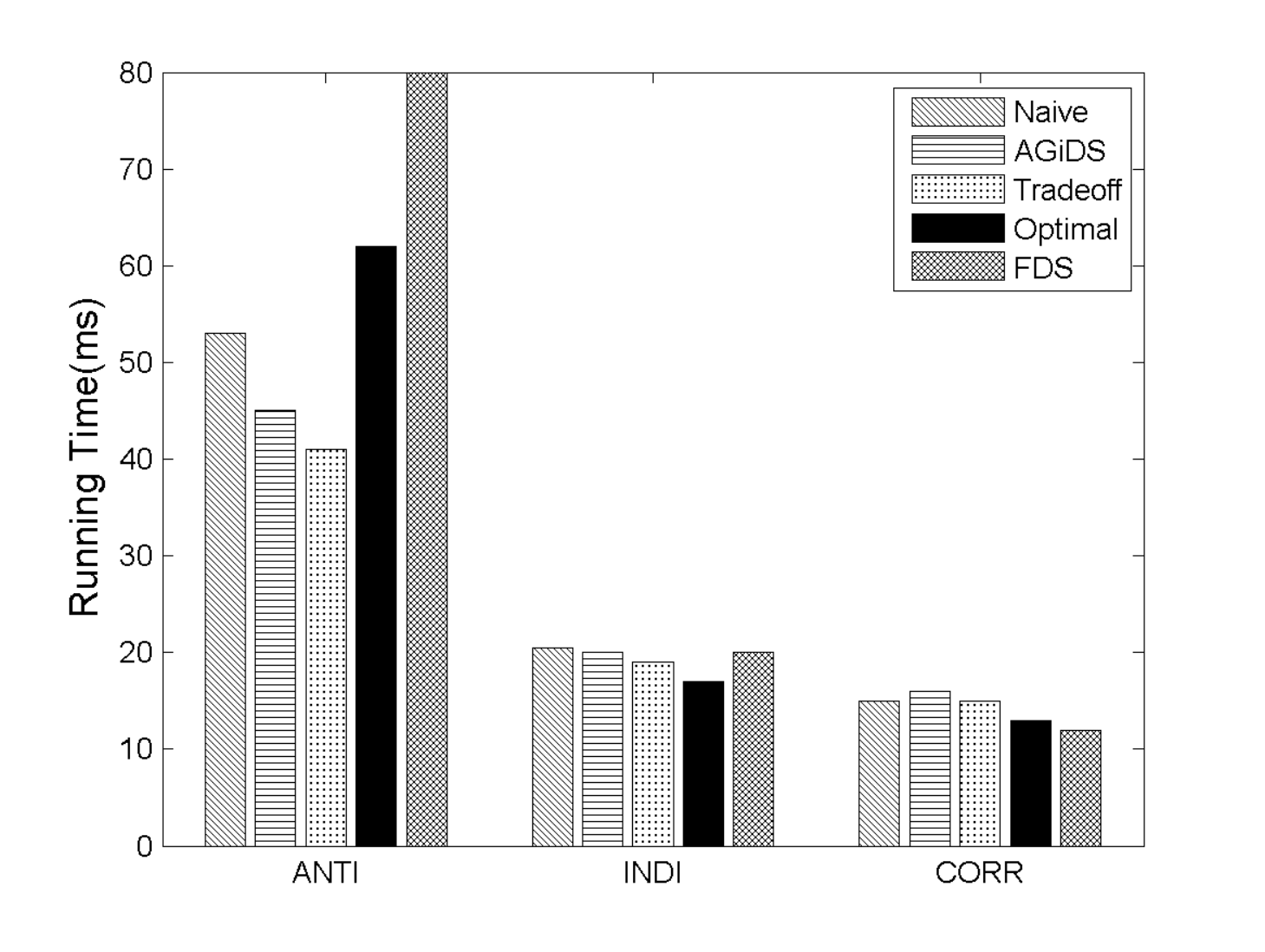}
\caption{Running time on synthetic datasets under horizontal partition (in \tradeoff\ we use $5$ rounds). The time for local skyline computation at the beginning is not included.}
\label{fig:hortime}
\end{figure}

\begin{figure}[t]
\centering
\includegraphics[height = 2in]{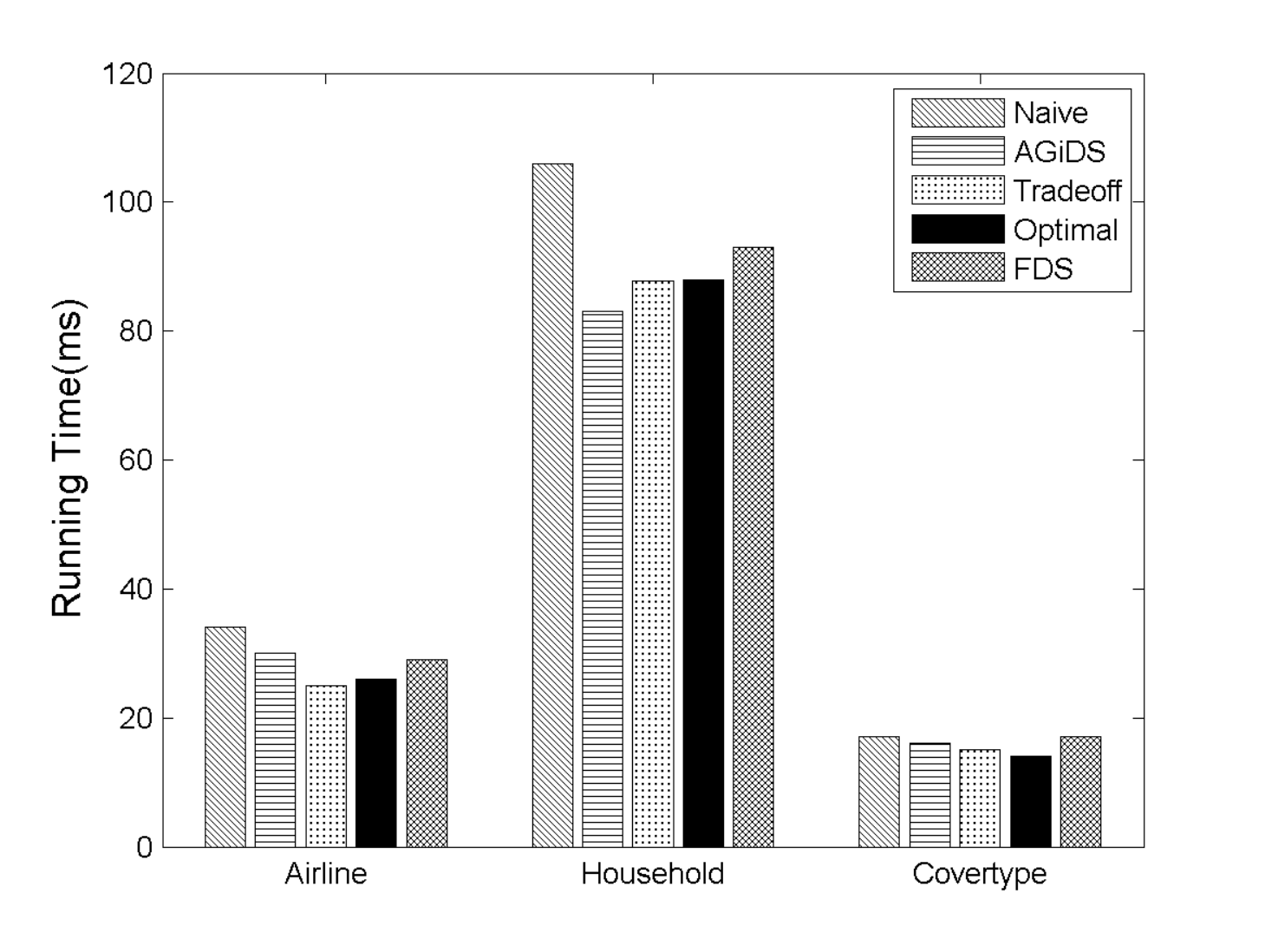}
\caption{Running time on real-world datasets under horizontal partition (in \tradeoff\ we use $5$ rounds). The time for local skyline computation at the beginning is not included.}
\label{fig:hortime1}
\end{figure}

Figure~\ref{fig:hortime} and Figure~\ref{fig:hortime1} show the running time of the tested algorithms (excluding the cost the common local skyline computation at the beginning) on synthetic and real-world datasets respectively.\footnote{The local skyline computation takes about $6$ seconds on all the three synthetic datasets and $1.1$, $0.6$, $0.2$ seconds on the \air, \house\ and \cover\ respectively. This cost in fact dominates the other time costs.}   Generally speaking the running time of all algorithms are similar.  On the \anti\ dataset, the running time of \opt\ and \fds\ are clearly worse than others.  This is because they need many more rounds than other algorithms on \anti.

\paragraph{Summary} \tradeoff\ achieves noticeable communication cost reductions than \agids\ and \naive\ by using one or two more rounds; its performance is very close to the theoretical optimal algorithm \opt\ in the communication cost but is much more efficient in rounds.  On the other hand, \agids\ does not have an advantage against \naive\ in communication but uses one more round, and the performance of \fds\ is clearly dominated by \tradeoff\ and \opt.  All algorithms have similar time costs since the computation of the local skylines dominates the other costs.

\subsection{Vertical Partition}
\label{sec:exp-ver}

We compare the following algorithms in the vertical partition case. 
\begin{itemize}
\item \prune:  Algorithm~\ref{alg:vertical} in Section~\ref{sec:heuristic}.

\item \bds\ and \ids: We use the \bds\ and \ids\  algorithms proposed in \cite{BGZ04} as comparisons.  See Section~\ref{sec:heuristic} for a brief description of these two algorithms.

\item \pds: We use the \pds\  algorithm proposed in \cite{LYLC06} as a comparison. See Section~\ref{sec:heuristic} for a brief description of this algorithm.
\end{itemize}

\begin{figure*}[t]
\begin{minipage}[d]{0.33\linewidth}
\centering
\includegraphics[width=1\textwidth]{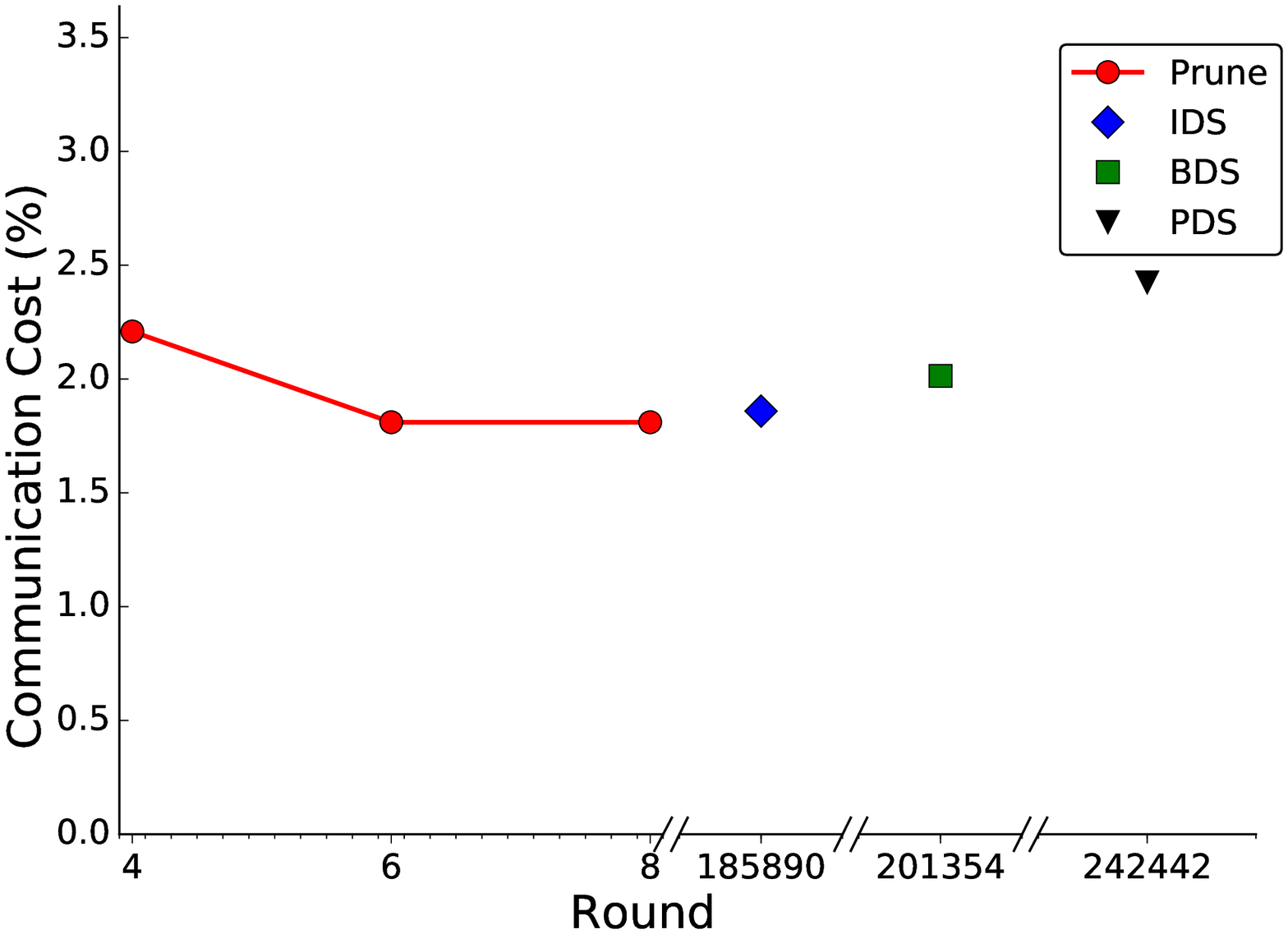}
\centerline{\anti}
\end{minipage}
\begin{minipage}[d]{0.33\linewidth}
\centering
\includegraphics[width=1\textwidth]{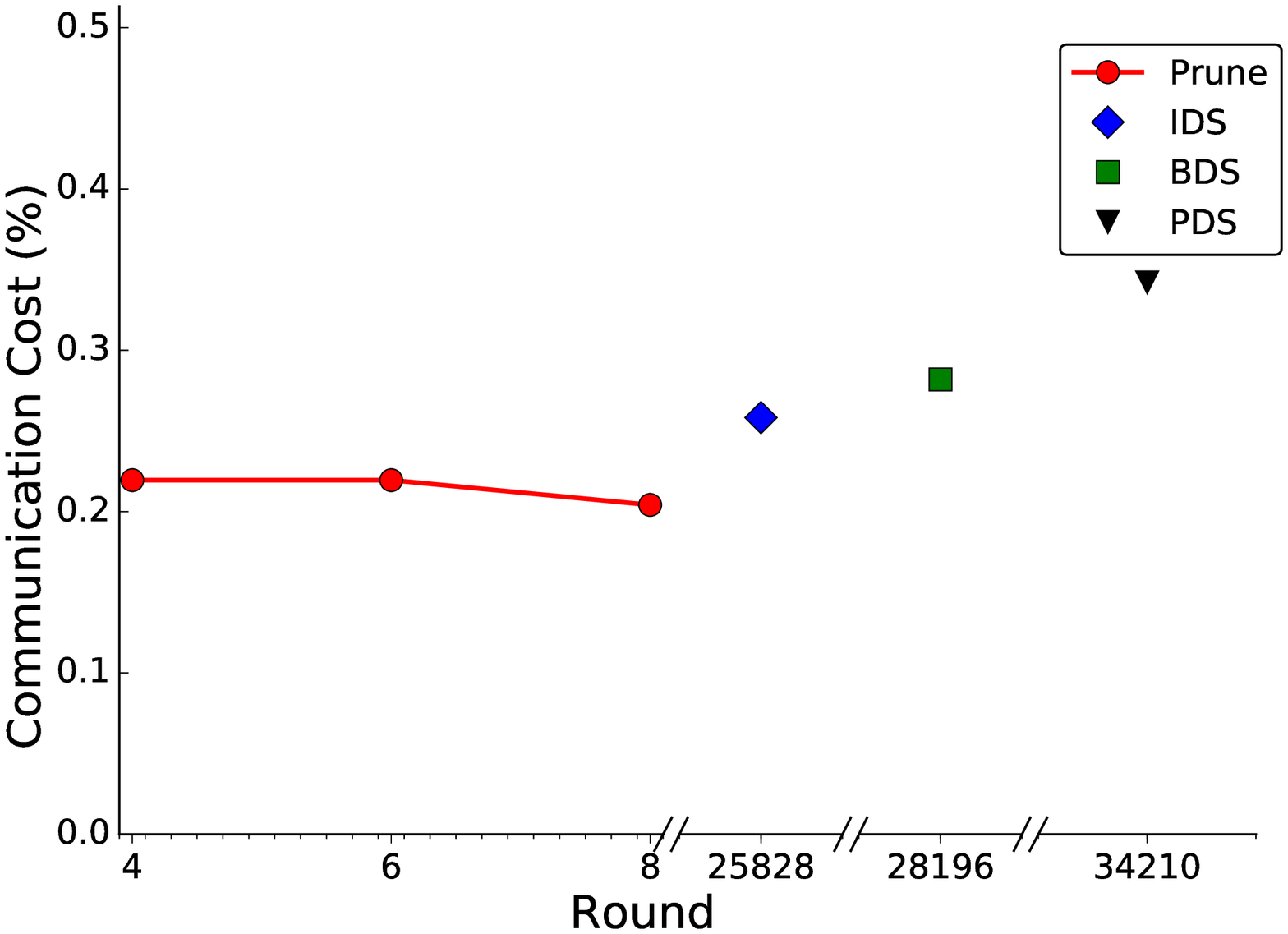}
\centerline{\ind}
\end{minipage}
\begin{minipage}[d]{0.33\linewidth}
\centering
\includegraphics[width=1\textwidth]{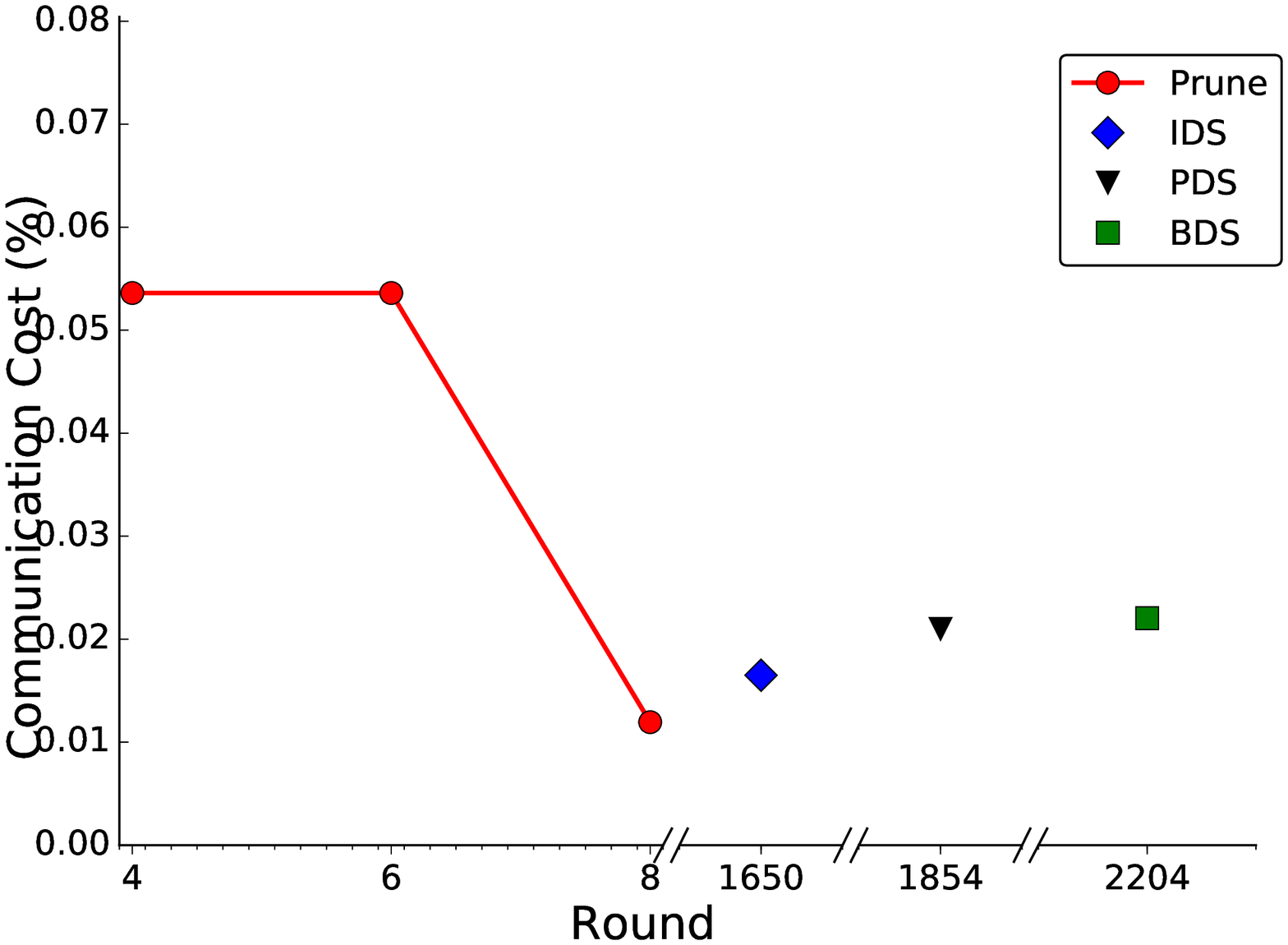}
\centerline{\cor}
\end{minipage}
\caption{Communication and round costs on synthetic datasets under vertical partition; for \prune\ we use parameter $\rho = 1000, 10000, 100000$ for \anti, \ind, \cor\ datasets respectively)}
\label{fig:verrsyn}
\end{figure*}

\begin{figure*}[t]
\begin{minipage}[d]{0.33\linewidth}
\centering
\includegraphics[width=1\textwidth]{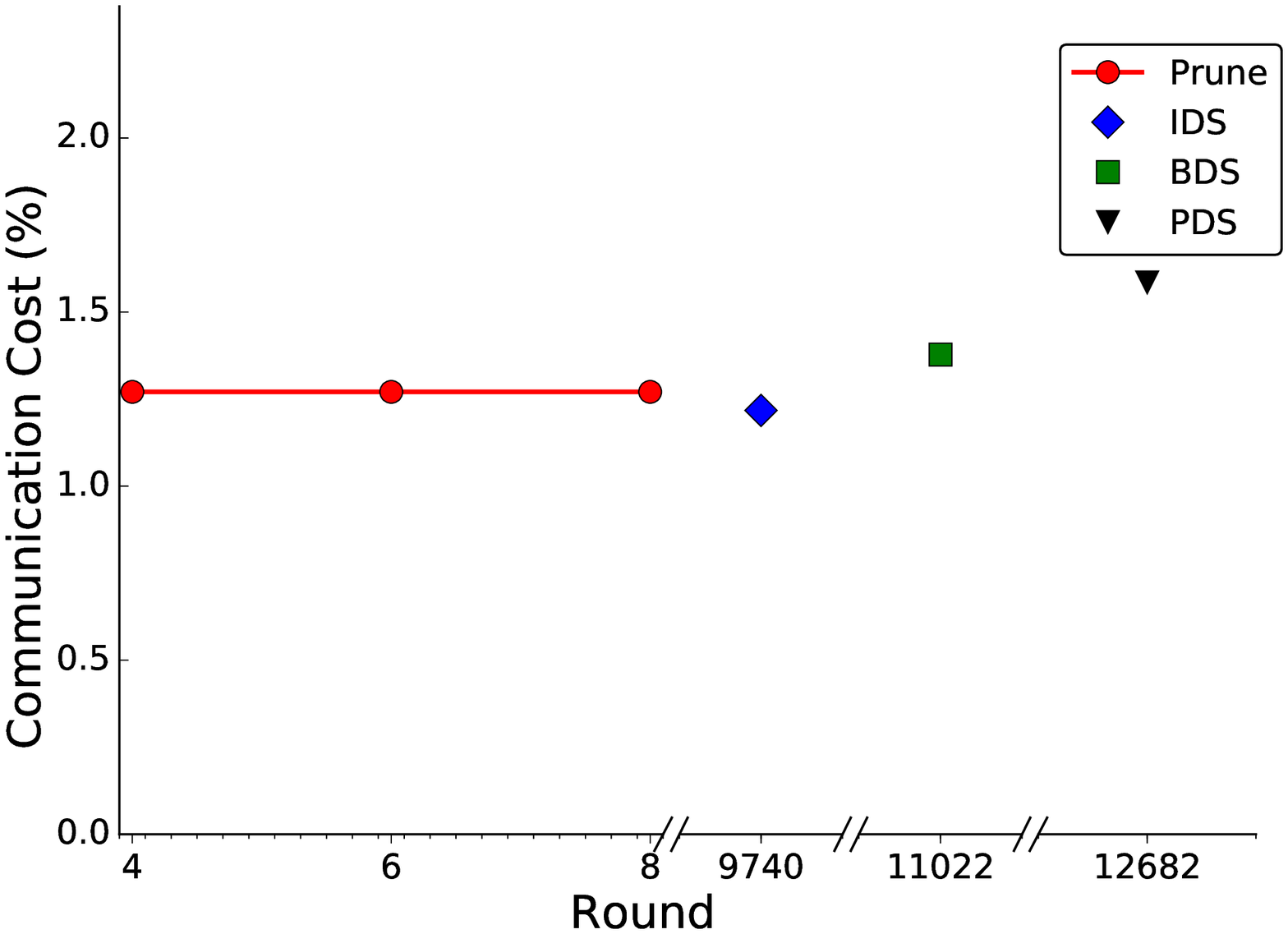}
\centerline{\air}
\end{minipage}
\begin{minipage}[d]{0.33\linewidth}
\centering
\includegraphics[width=1\textwidth]{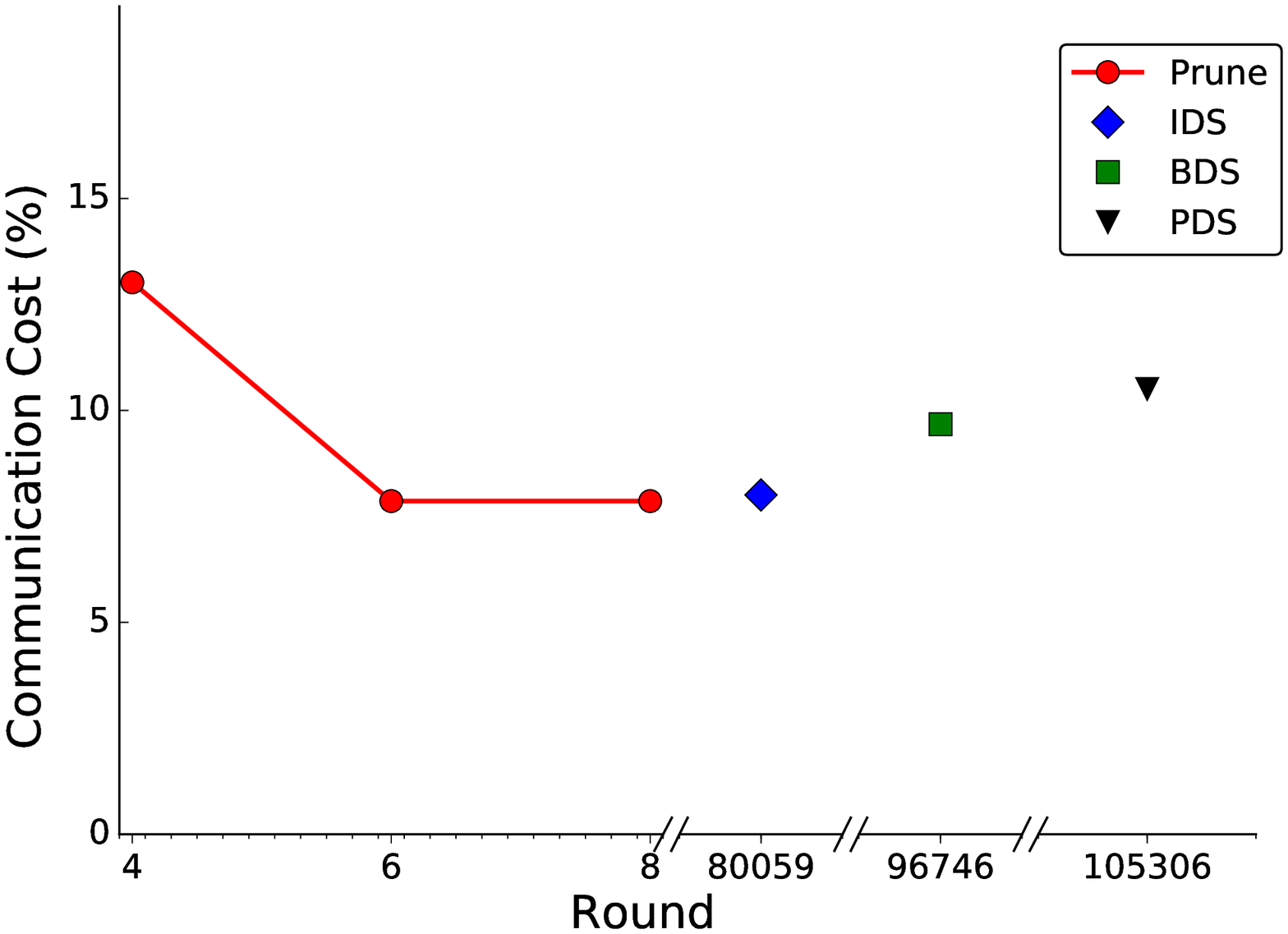}
\centerline{\house}
\end{minipage}
\begin{minipage}[d]{0.33\linewidth}
\centering
\includegraphics[width=1\textwidth]{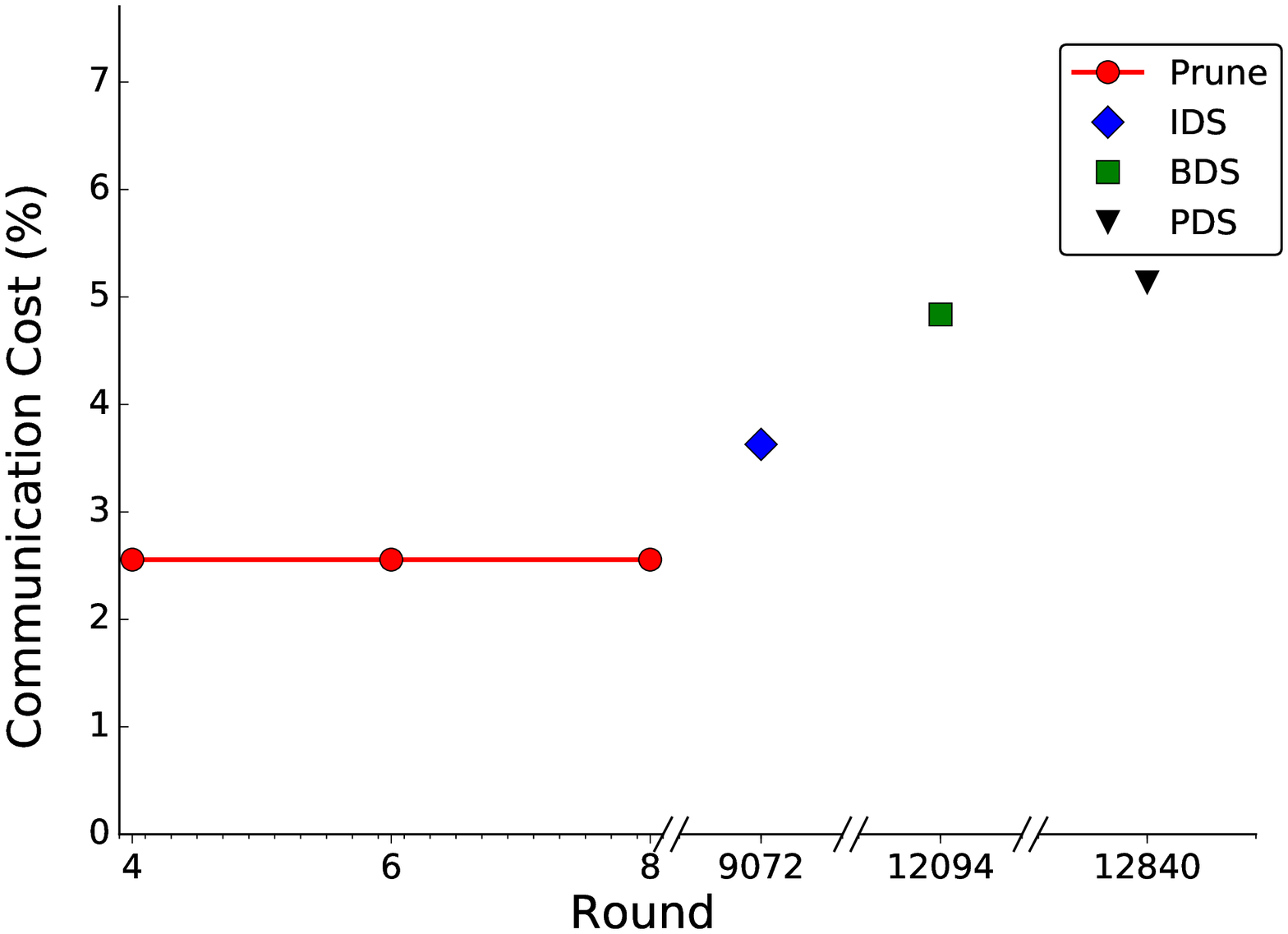}
\centerline{\cover}
\end{minipage}
\caption{Communication and round costs on real-world datasets under vertical partition; for \prune\ we use parameter $\rho = 500$ for all \air, \house, \cover\ datasets)}
\label{fig:verrreal}
\end{figure*}

Figure~\ref{fig:verrsyn} and Figure~\ref{fig:verrreal} show the communication and round costs of the four tested algorithms on the three synthetic datasets and three real-world datasets mentioned above.  We observed that by using at most eight rounds, the communication cost of \prune\ is similar or even smaller than that of \bds, \ids\ and \pds, which use significantly more rounds.   It is surprising to see that \pds\ is even worse then \bds\ in most cases; the original paper \cite{LYLC06} did not make this comparison on points in the 2D Euclidean space.  One reason that \prune\ can achieve better communication costs is that \ids\ and \pds\ have significantly larger round complexities, and at every two rounds the coordinator needs to send a message to a site to request for the next point, which can be considered as an overhead of the algorithms in the distributed setting.  We also observe that \prune\ recovers similar number of points as \bds, \ids, and \pds; see Table~\ref{tab:points} for the details.

\begin{table}[t]
\centering
\begin{tabular}{|c|l|l|l|l|} \hline
&\prune & \ids & \bds & \pds \\ \hline
\anti &136094 &92945 &100677 &121221\\ \hline
\ind &13609 &12914 &14098 &17105\\ \hline
\cor &855 &825 &1102 &927\\ \hline
\air &6776 &4870 &5511 &6341\\ \hline
\house &52407 &40029 &48373 &52653\\ \hline
\cover &5258 &4536 &6047 &6420\\ \hline
\end{tabular}
\caption{Number of recovered points at the coordinator. In \prune, $\rho$ is chosen to be $1000$, $10000$, $100000$ for \anti, \ind\ and \cor\ respectively, and $500$ for \air, \house\ and \cover; the round budget $r$ is set to be $8$.}
\label{tab:points}
\end{table}

In Figure~\ref{fig:verrsyn} and Figure~\ref{fig:verrreal} we have set $\rho$ to be $1000$, $10000$, $100000$ for three synthetic datasets \anti, \ind\ and \cor, and $500$ for all three real-world datasets. These values are not chosen optimally for each dataset, but follow some general rules we have concluded from another set of experiments, which we discuss next.

\begin{figure*}[t]
\begin{minipage}[d]{0.33\linewidth}
\centering
\includegraphics[width=1\textwidth]{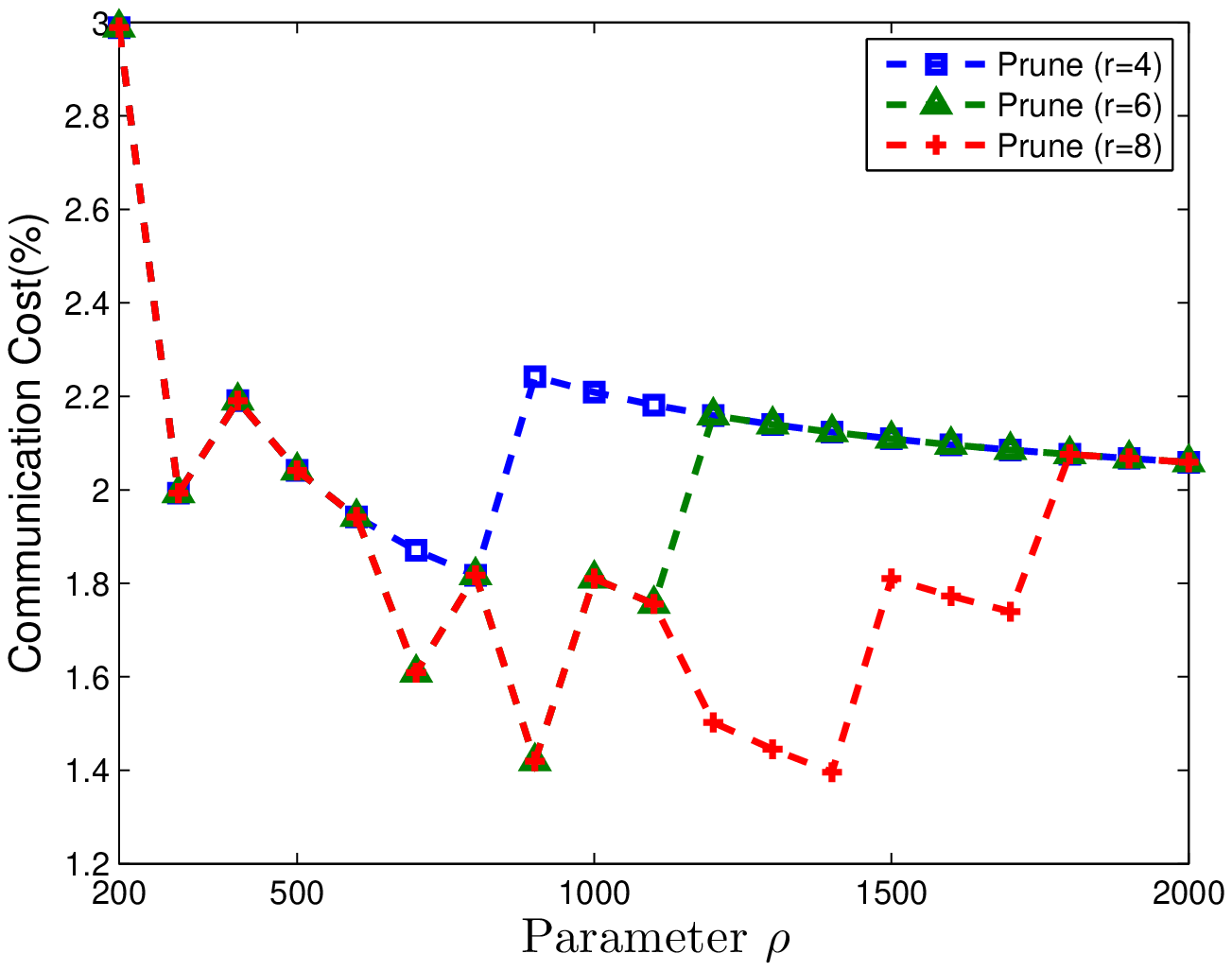}
\centerline{\anti}
\end{minipage}
\begin{minipage}[d]{0.33\linewidth}
\centering
\includegraphics[width=1\textwidth]{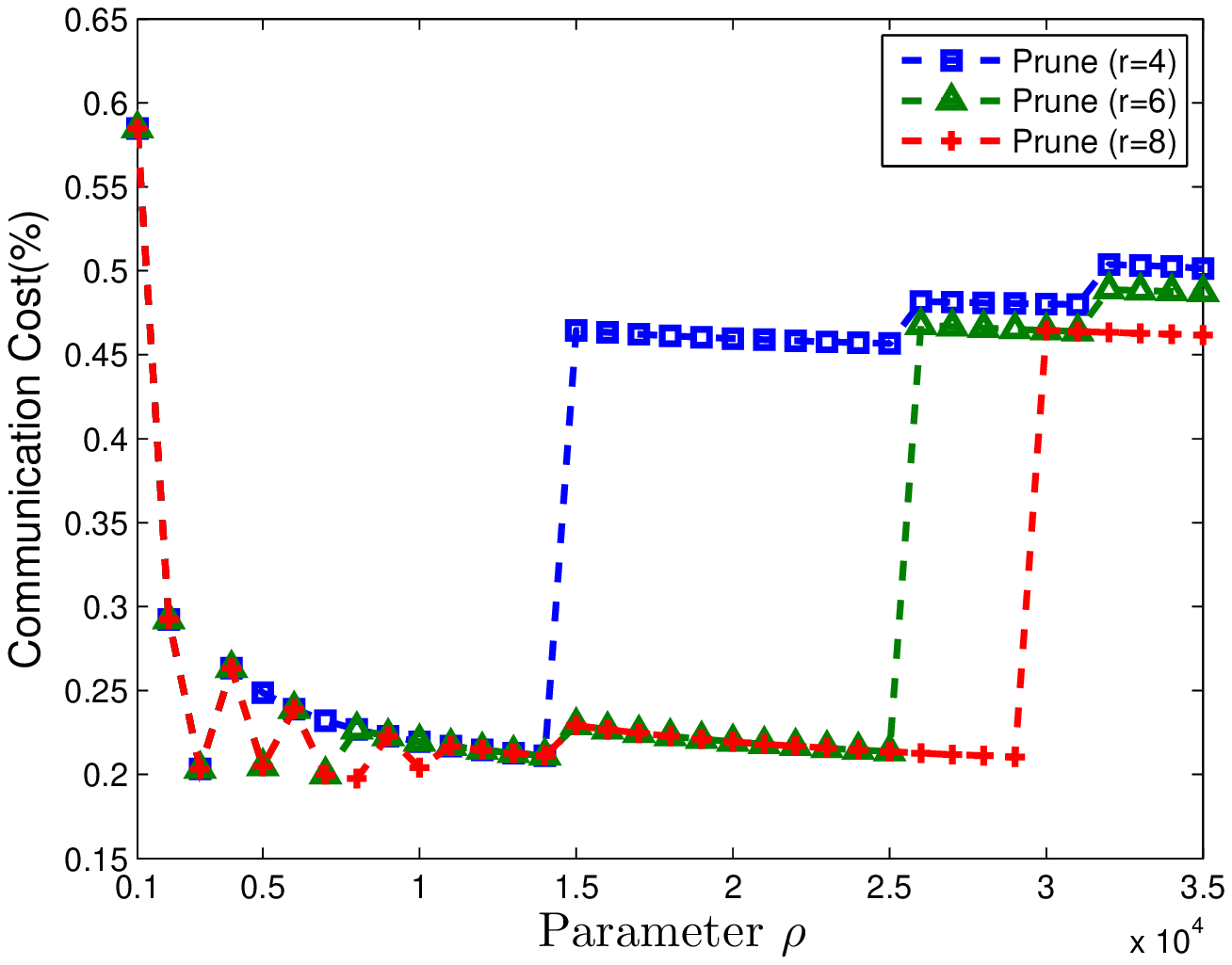}
\centerline{\ind}
\end{minipage}
\begin{minipage}[d]{0.33\linewidth}
\centering
\includegraphics[width=1\textwidth]{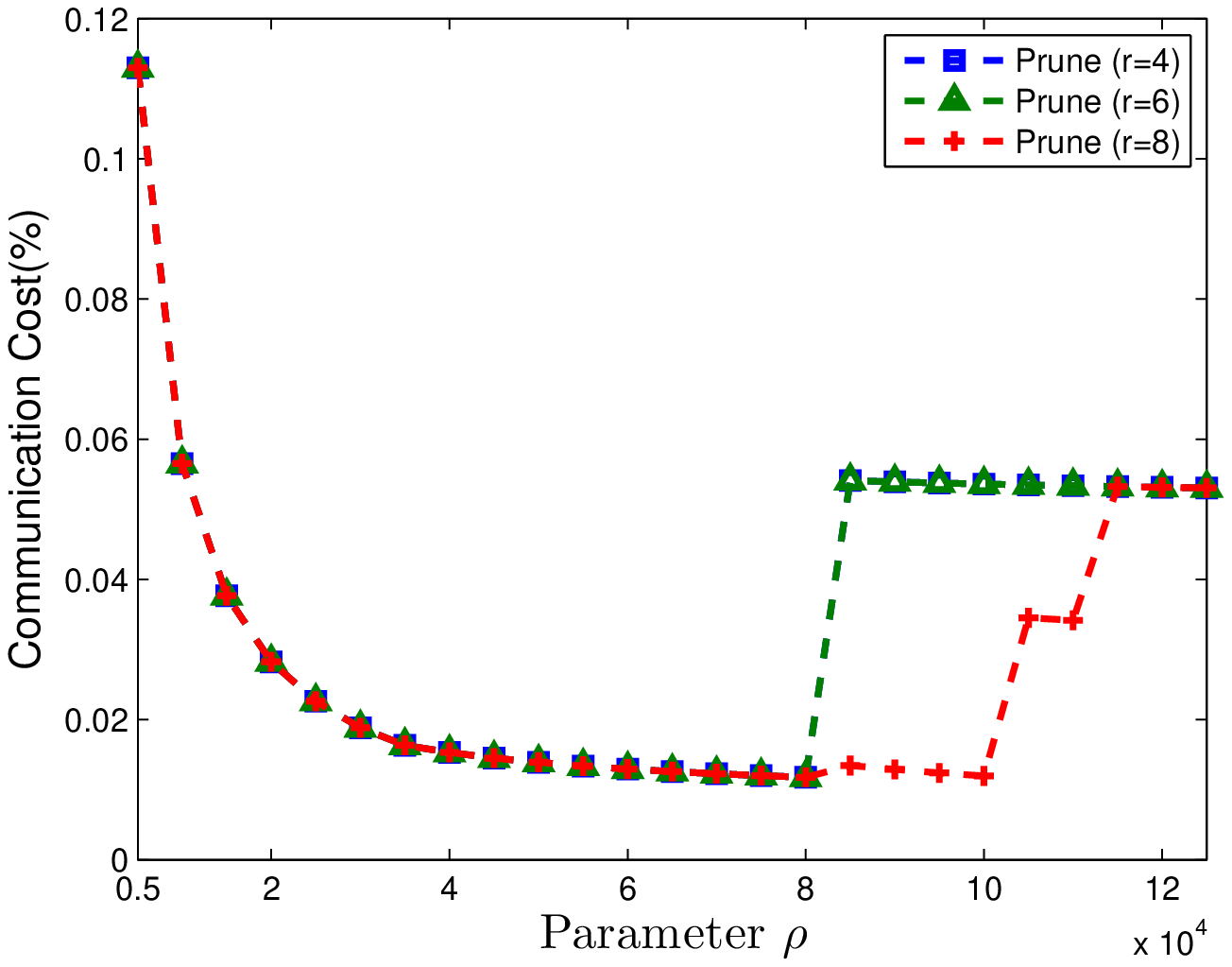}
\centerline{\cor}
\end{minipage}
\caption{Communication cost versus parameter $\rho$ on synthetic dataset under vertical partition}
\label{fig:verdcostsys}
\end{figure*}

\begin{figure*}[t]
\begin{minipage}[d]{0.33\linewidth}
\centering
\includegraphics[width=1\textwidth]{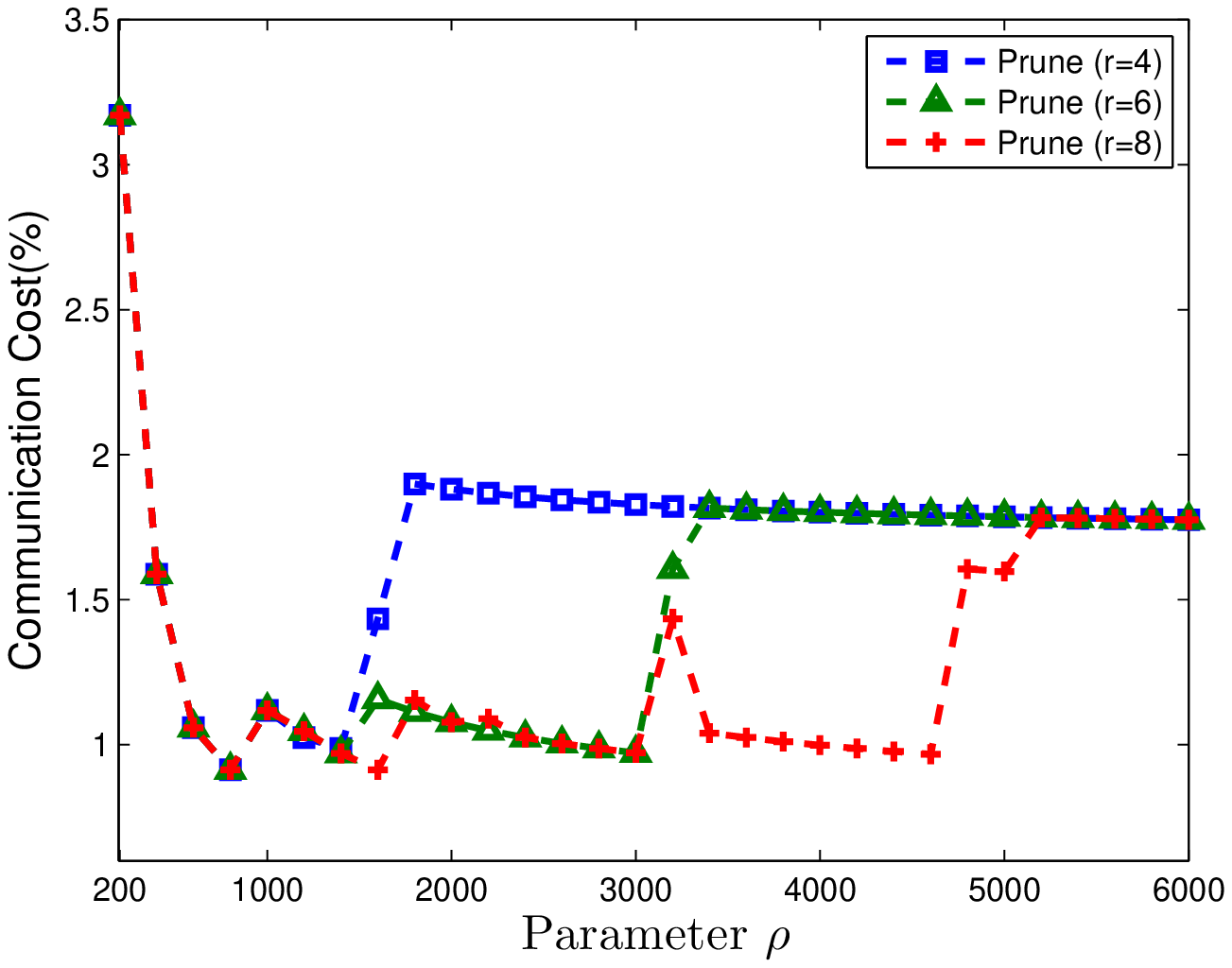}
\centerline{\air}
\end{minipage}
\begin{minipage}[d]{0.33\linewidth}
\centering
\includegraphics[width=1\textwidth]{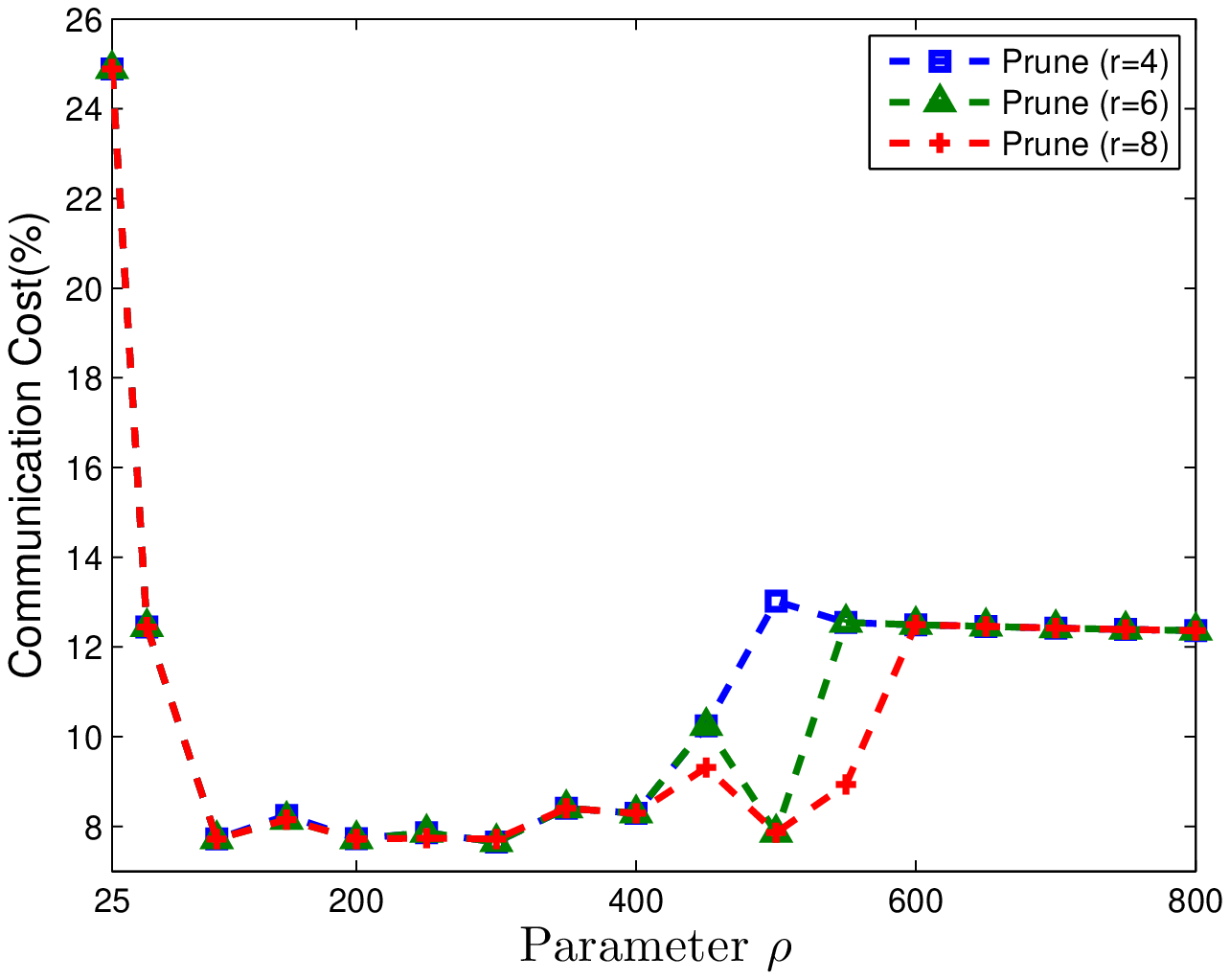}
\centerline{\house}
\end{minipage}
\begin{minipage}[d]{0.33\linewidth}
\centering
\includegraphics[width=1\textwidth]{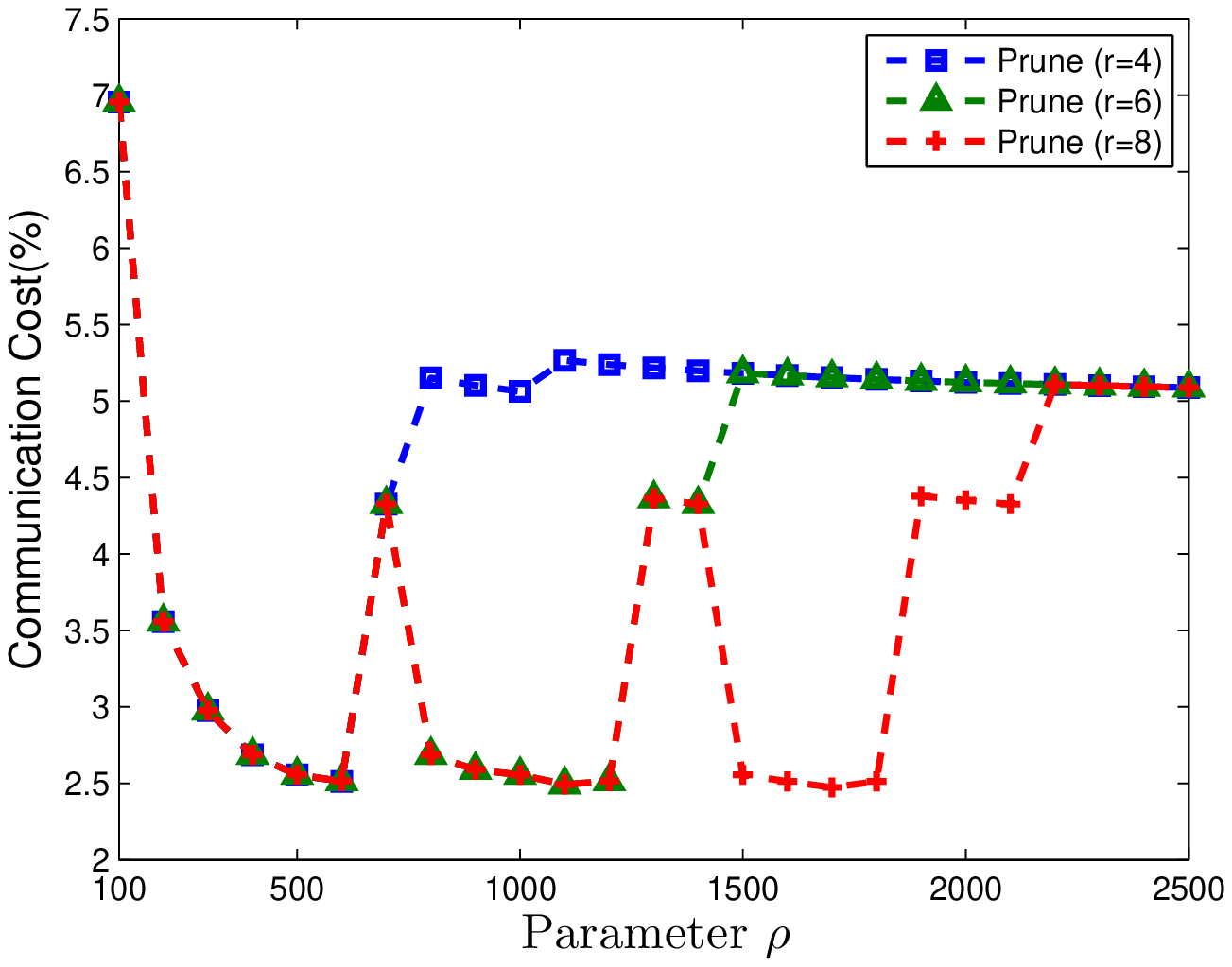}
\centerline{\cover}
\end{minipage}
\caption{Communication cost versus parameter $\rho$ on synthetic dataset under vertical partition}
\label{fig:verdcostreal}
\end{figure*}

Figure~\ref{fig:verdcostsys} and Figure~\ref{fig:verdcostreal} show how the communication cost changes when varying $\rho$.
We observe that for a fixed round budget, the communication cost curves generally obey a ``U'' shape: the cost is very large when $\rho$ is very small; this is because the group size is too large so many recovered points are wasted in the first few rounds.  The cost then deceases dramatically as $\rho$ increases, becomes stable for a while, and then increases again.  We also observe that the stable range of $\rho$ becomes larger when the round budget $r$ becomes larger, which makes the choice of a good $\rho$ more flexible.  While the best $\rho$ inherently depends on the distribution of the dataset, the lesson we have learned from Figure~\ref{fig:verdcostsys} and Figure~\ref{fig:verdcostreal} is that $\rho$ should be small (e.g., $1000$) if the data distribution is close to \anti, large (i.e., $100000$) if the distribution is close to \cor, and medium (i.e., 10000) if the distribution is close to \ind.  It is reasonable to believe that most real-world datasets are on the \anti\ end, since an item which is strong on one attribute may be weak on the other (think about the price and location of hotels).  Our experiments on three real-world datasets \air, \house and \cover\ show that $\rho = 500$ can be a good default choice in practice.

\begin{figure}[t]
\centering
\includegraphics[height = 2in]{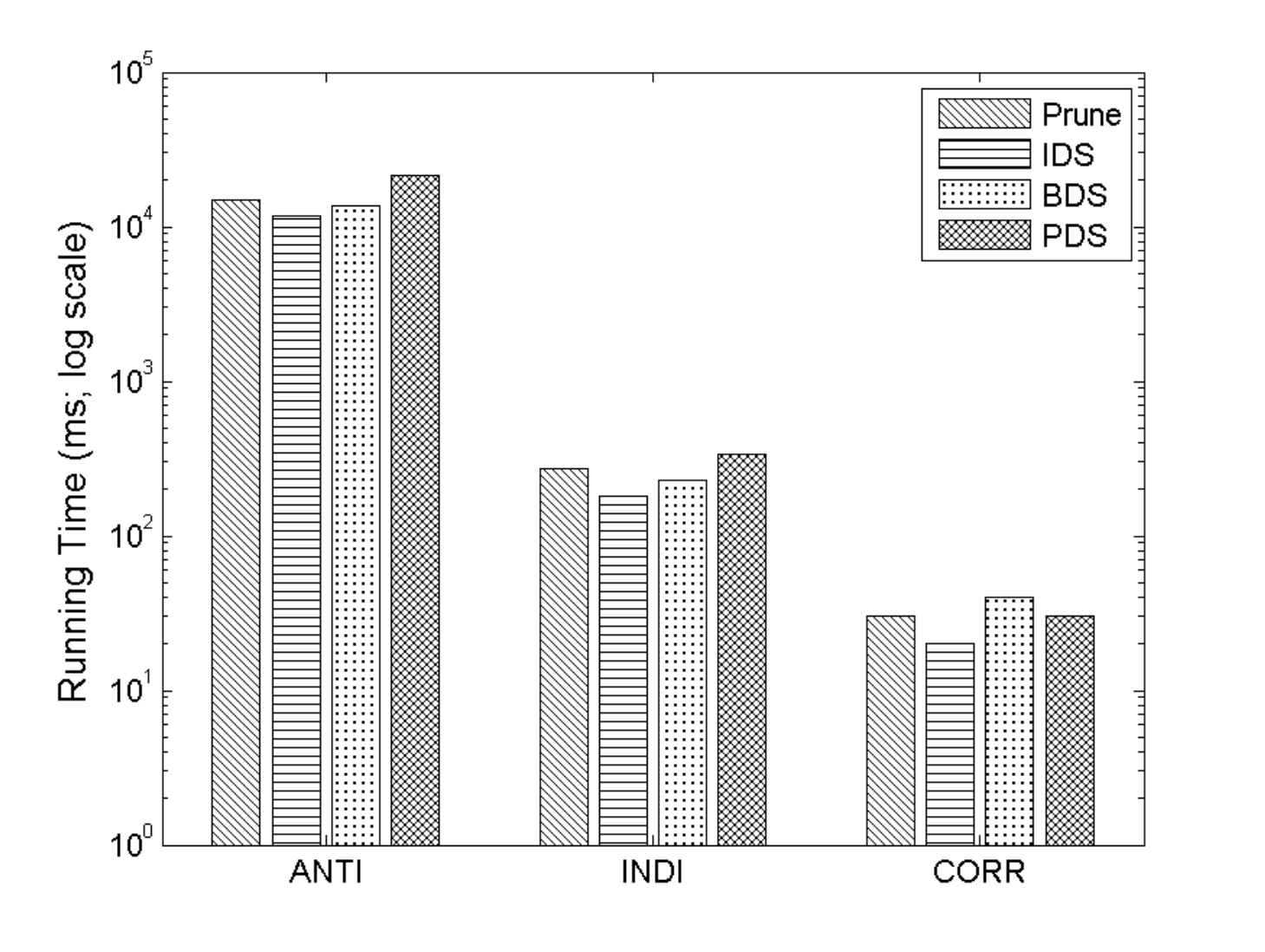}
\caption{Running time on synthetic datasets under vertical partition; for \prune\ we use parameter $\rho=1000, 10000, 100000$ for \anti, \ind, \cor\ datasets respectively, and $r = 8$ rounds). The time for local skyline computation at the beginning is not included.}
\label{fig:vertime}
\end{figure}

\begin{figure}[t]
\centering
\includegraphics[height = 2in]{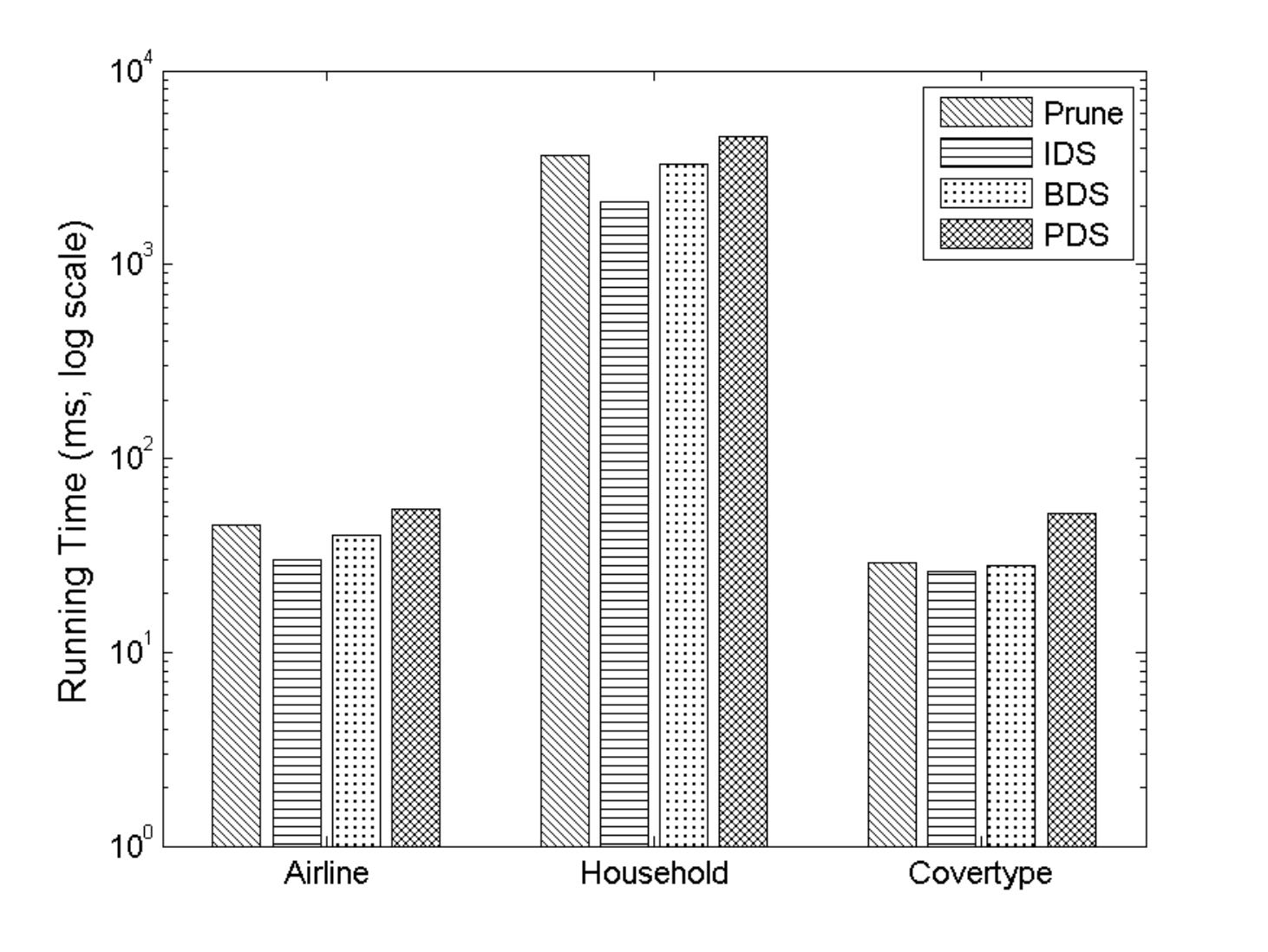}
\caption{Running time on real-world datasets under vertical partition; for \prune\ we use parameter $\rho = 500$ for all \air, \house, \cover\ datasets, and $r = 8$ rounds). The time for local skyline computation at the beginning is not included.}
\label{fig:vertime1}
\end{figure}

Figure~\ref{fig:vertime} and Figure~\ref{fig:vertime1} show the time costs of tested algorithms on synthetic and real-world datasets respectively.  We do not take into account the initial sorting time which is common to all algorithms.\footnote{The sorting time is about $5$ seconds on all the three synthetic datasets, and $0.9$, $0.5$, $0.15$ seconds on \air, \house\ and \cover.}
Generally speaking, the running times of all algorithms are similar.  This is because all of the three algorithms are based on the TA approach.  \pds\ has the worst time performance in most cases because it needs to recover the largest number of points {\em and} uses the most number of rounds.  

\paragraph{Summary} To conclude, \prune\ achieves a significant reduction on the round costs (3-4 orders of magnitude) while preserves  communication costs compared with \ids, which is the best among the three existing algorithms.  The running times of all the tested algorithms are similar.

\section{Conclusion}
\begin{figure*}[]
\begin{minipage}[d]{0.33\linewidth}
\centering
\includegraphics[width=0.6\textwidth]{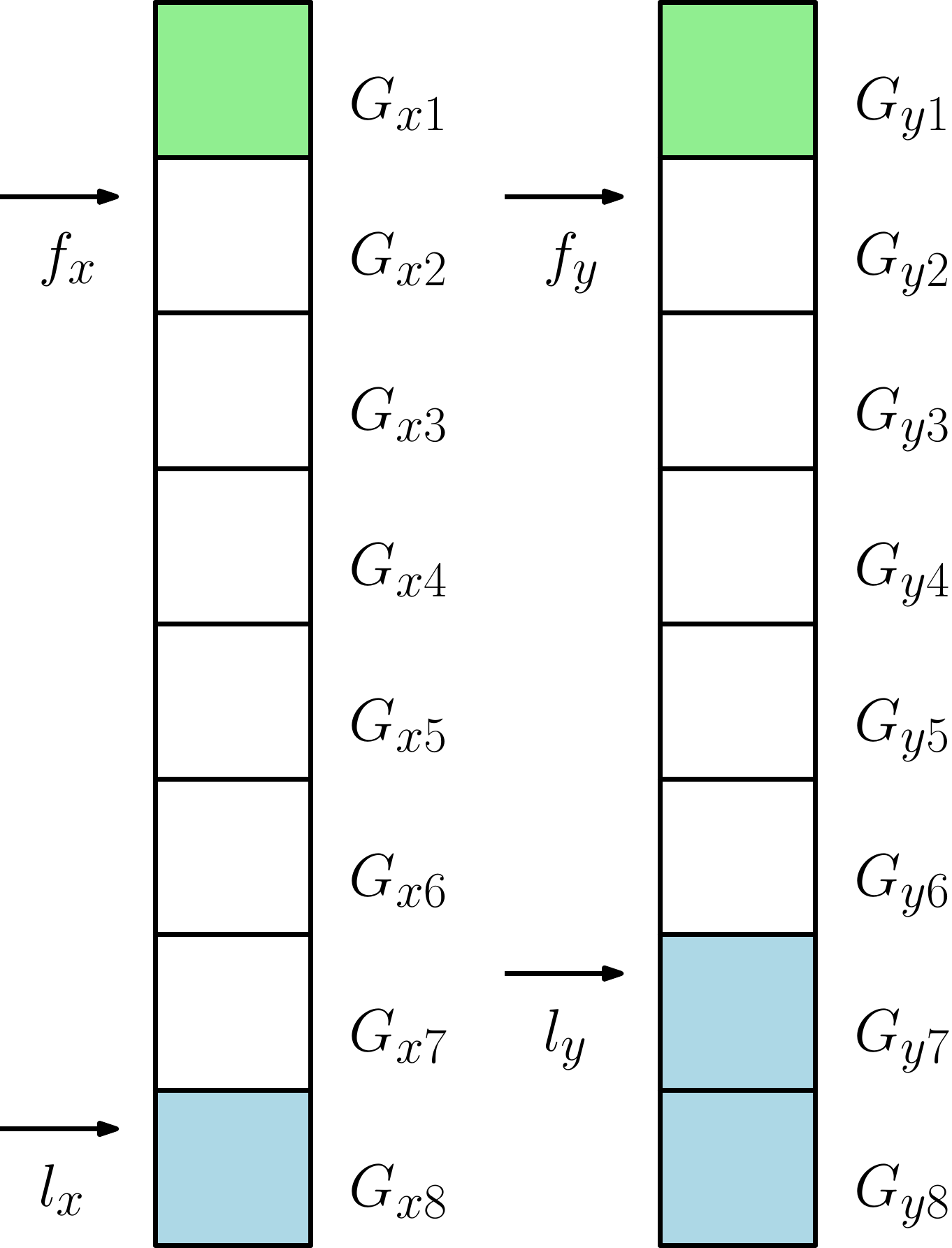}
\centerline{$Round=1,2$}
\end{minipage}
\begin{minipage}[d]{0.33\linewidth}
\centering
\includegraphics[width=0.6\textwidth]{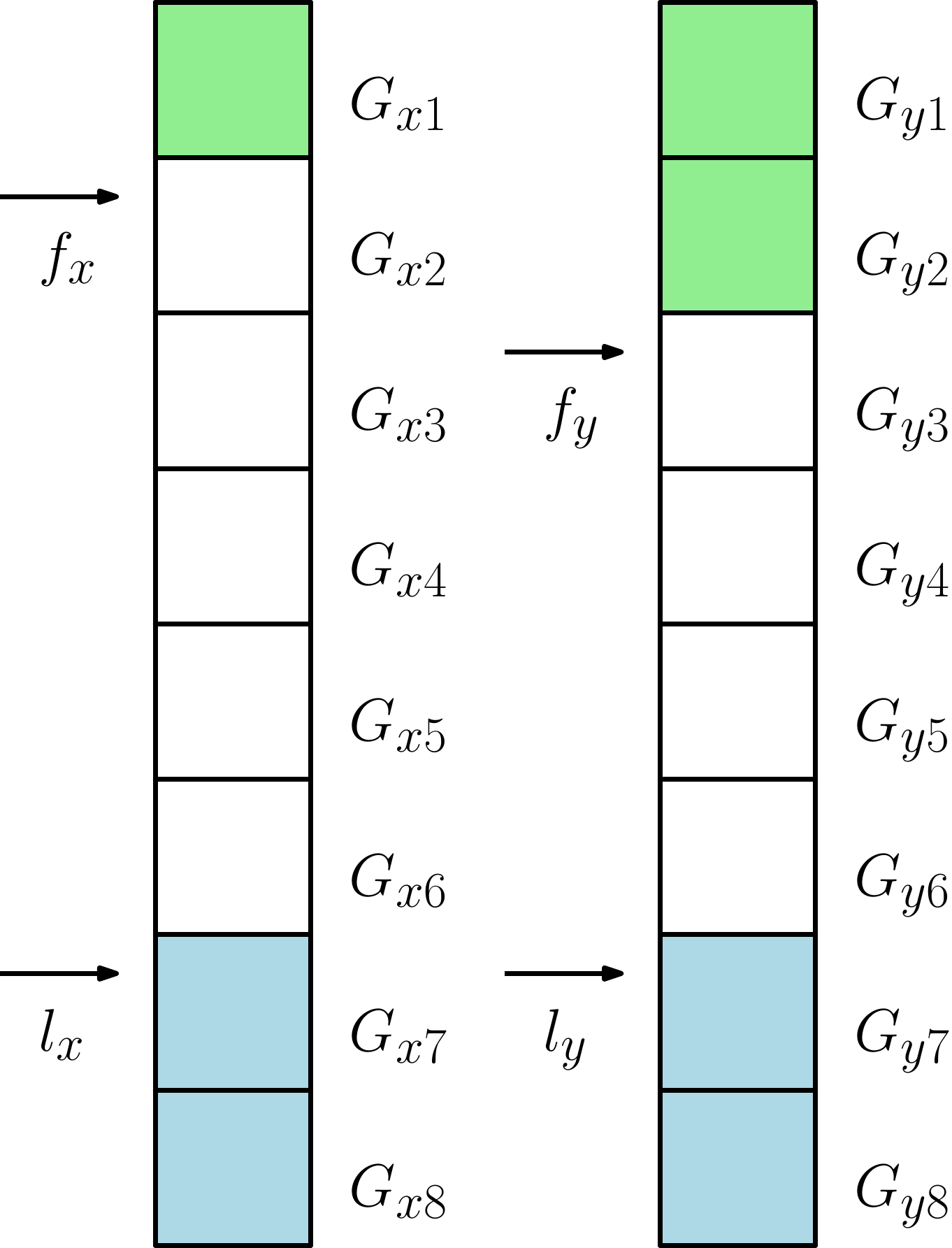}
\centerline{$Round=3,4$}
\end{minipage}
\begin{minipage}[d]{0.33\linewidth}
\centering
\includegraphics[width=0.6\textwidth]{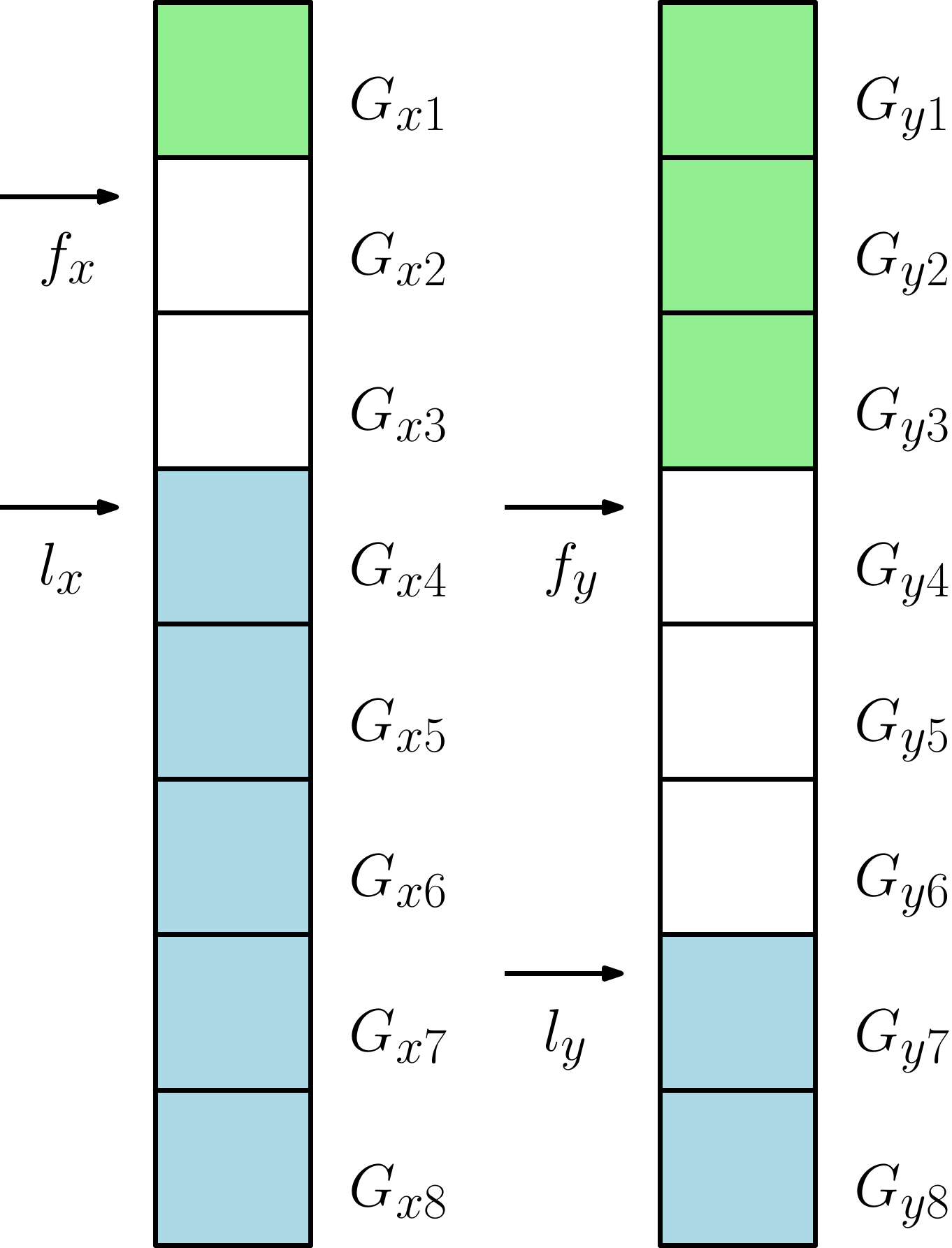}
\centerline{$Round=5,6$}
\end{minipage}
\caption{A running example of Algorithm~\ref{alg:vertical}, with parameters $r=8$ and $\rho=8$.}
\label{fig:vertical}
\end{figure*}
In this paper we propose a set of algorithms for computing skylines on distributed data.  We consider both horizontal and vertical data partitions. For horizontal partition we give an algorithm that achieves the optimal communication cost.  We also propose an algorithm with a smooth communication-round tradeoff.  We show experimentally that our algorithms (with theoretical guarantees) significantly outperform existing heuristics in the communication cost and/or round cost.   For vertical partition, we first show a strong lower bound, and then give a practical heuristic. Our heuristic significantly outperforms existing ones in the round cost, while achieves similar or better communication cost.  Further work includes designing algorithms for points in the higher dimensional Euclidean spaces for both horizontal and vertical data partitions.

\bibliography{paper}
\bibliographystyle{abbrv}

\appendix
\balance
\section{A Running Example for the Interactive Pruning Algorithm}
\label{sec:example}

Figure~\ref{fig:vertical} shows a running example of Algorithm~\ref{alg:vertical} with parameters $r=8$ and $\rho=8$. In the first round, Alice sends $G_{x1}$ and Bob sends $G_{y1}$ to the coordinator. In the second round,  the coordinator sends the IDs of points in $G_{x1}$ and $G_{y1}$ to Bob and Alice respectively, and then Bob and Alice send the $y$-coordinates and $x$-coordinates of points in $G_{x1}$ and $G_{y1}$ respectively to the coordinator. The coordinator recovers the points in $G_{x1}$ and $G_{y1}$ and adds them to $R$ (the set of candidate skyline points), and updates $l_x = 8$, $l_y = 7$, and $f_x = f_y = 2$.  In the third and fourth rounds,  the coordinator chooses to recover points in $G_{y2}$ because $l_y - f_y < l_x - f_x$, and then updates $l_x = 7$. In the fifth and sixth rounds,  the coordinator chooses to recovers points in $G_{y3}$ because $l_y - f_y < l_x - f_x$, and then updates $l_x = 4$. In the last two rounds, the coordinator asks Alice for the information of the points in the remaining groups $G_{x2}$ and $G_{x3}$ (i.e., the groups that have {\em not} been recovered or pruned) since $l_x - f_x < l_y - f_y$.

\end{document}